\RequirePackage{fix-cm}
\documentclass[twocolumn,epjc3]{svjour3}  
\smartqed  
\RequirePackage{graphicx}
\journalname{Eur. Phys. J. C}
\usepackage{amssymb}
\usepackage{amsmath}
\usepackage{amsfonts}
\usepackage{slashed}
\usepackage{enumerate}
\usepackage{graphicx}
\usepackage{relsize}
\usepackage{color}
\usepackage{extarrows}
\usepackage{cancel}
\usepackage{placeins}
\usepackage{float}
\usepackage{subcaption}
\usepackage{caption}
\usepackage{here}
\usepackage{cite}
\definecolor{gray}{rgb}{0.5, 0.5, 0.5}
\usepackage{xcolor,soul}
\sethlcolor{yellow}
\RequirePackage{chapterbib}

\usepackage{hyperref}
\hypersetup{
  colorlinks=true,        
  linkcolor=blue,         
  citecolor=cyan,         
}

\begin{document}

\title{Probing Geometric Proca in Metric-Palatini Gravity with Black Hole Shadow and Photon Motion}


\author{Elham Ghorani$ \thanksref{addr1,e1}$
        \and
        Beyhan Puli{\c c}e$ \thanksref{addr1,e2}$
        \and
        Farruh Atamurotov$  \thanksref{addr2,addr3,addr4,addr5,e3}$
        \and
        Javlon~Rayimbaev$ \thanksref{addr6,addr7,addr4,addr8,addr9,e4}$
        \and
        Ahmadjon Abdujabbarov$ \thanksref{addr10,addr4,addr8,e5}$
        \and
        Durmu{\c s} Demir$\thanksref{addr1,e6}$
}

\thankstext{e1}{e-mail: elham.ghorani@sabanciuniv.edu}
\thankstext{e2}{e-mail: beyhan.pulice@sabanciuniv.edu}
\thankstext{e3}{e-mail: atamurotov@yahoo.com}
\thankstext{e4}{e-mail: javlon@astrin.uz}
\thankstext{e5}{e-mail: ahmadjon@astrin.uz}
\thankstext{e6}{e-mail: durmus.demir@sabanciuniv.edu}

\institute{Faculty of Engineering and Natural Sciences, Sabanc{\i} University, 34956 Tuzla, {\.Istanbul}, Turkey \label{addr1}
\and
New Uzbekistan University, Mustaqillik ave. 54, 100007 Tashkent, Uzbekistan \label{addr2}
\and
Inha University in Tashkent, Ziyolilar 9, Tashkent 100170, Uzbekistan \label{addr3}
\and
National University of Uzbekistan, Tashkent 100174, Uzbekistan \label{addr4}
\and
University of Public Safety of the Republic of Uzbekistan, Tashkent Region 100109, Uzbekistan \label{addr5}
\and
Institute of Fundamental and Applied Research, National Research University TIIAME, Kori Niyoziy 39, Tashkent 100000, Uzbekistan \label{addr6}
\and
Akfa University, Milliy Bog' Street 264, Tashkent 111221, Uzbekistan \label{addr7}
\and
Tashkent State Technical University, Tashkent 100095, Uzbekistan\label{addr8}
\and 
Samarkand State University, University Avenue 15, Samarkand 140104, Uzbekistan \label{addr9}
\and
Ulugh Beg Astronomical Institute, Astronomy St 33, Tashkent 100052, Uzbekistan \label{addr10}
}

\date{Received: date / Accepted: date}

\maketitle

\begin{abstract}
Extended metric-Palatini gravity, quadratic in the antisymmetric part of the affine curvature, is known to lead to the general relativity plus a geometric Proca field. The geometric Proca, equivalent of the non-metricity vector in the torsion-free affine connection, qualifies to be a distinctive signature of the affine curvature. In the present work, we explore how shadow and photon motion near black holes can be used to probe the geometric Proca field. To this end, we derive static spherically symmetric field equations of this Einstein-geometric Proca theory, and show that it admits black hole solutions in  asymptotically AdS background.  We perform a detailed study of the optical properties and shadow of this black hole and contrast them with the observational data by considering black hole environments with and without plasma. As a useful astrophysical application, we discuss constraints on the Proca field parameters using the observed angular size of the shadow of supermassive black holes M87$^*$ and Sgr A$^*$ in both vacuum and plasma cases.  Overall, we find that the geometric Proca can be probed via the black hole observations.

\end{abstract}

\section{Introduction}
In the present work, our goal is to probe a minimal form of Weyl gravity by the use of existing observational data on black holes. The work can be divided in two parts: The model and its black hole candidates (Sec. 2 and Sec. 3) and observational implications of the candidate black holes (Sec. 4, Sec. 5 and Sec. 6).

The question of if the general relativity (GR) is the sole theory of the gravitation is at the heart of the ongoing research in cosmology and astrophysics. In this regard, studying testable extensions of the GR proves particularly useful. One class of extensions concerns higher-order curvature invariants as in, for example, the $f(R)$ gravity \cite{f(R)-gravity}. One other class involves extension of the metrical geometry of the GR to non-Riemannian geometries based on metric-incompatible connections \cite{schroedinger,mag1,Vitagliano2011}. Our framework is a special case of such extensions.

The simplest non-Riemannian extension is the {\it Palatini formulation} \cite{Palatini}, which is characterized by metric $g_{\mu\nu}$ and the Ricci curvature ${\mathbb{R}}_{\mu\nu}(\Gamma)$ of a general symmetric affine connection $\Gamma^{\lambda}_{\mu\nu}$ (a torsion-free connection independent of the metric $g_{\mu\nu}$ and its Levi-Civita connection ${}^g\Gamma^{\lambda}_{\mu\nu}$). This formulation gives the Einstein field equations with no need to extrinsic curvature \cite{york,gh}. With general curvature invariants, it leads to the GR along with geometrical scalars,  vectors and tensors \cite{Demir2012}. The relevance of the Palatini gravity for emergent gravity theories \cite{Demir2019, Demir2021} is it's another application area. It was shown that the extension of the Palatini gravity with fundamental scalars like the Higgs field leads to natural inflation \cite{bauer-demir1,bauer-demir2}. Higher-curvature terms were also studied in the Palatini formalism \cite{Vitagliano2011,Vitagliano2013,Demir2020} and their certain effects in astrophysics and cosmology were analysed in \cite{Palatini-f(R)}.

One step further from the Palatini formulation is the inclusion of a term like ${\mathbb{R}}_{[\mu\nu]}(\Gamma) {\mathbb{R}}^{[\mu\nu]}(\Gamma)$, where ${\mathbb{R}}_{[\mu\nu]}(\Gamma)$ is the anti-symmetric part of the affine Ricci tensor ${\mathbb{R}}_{\mu\nu}(\Gamma)$ . What is important about this inclusion is that it leads dynamically to the GR plus a  purely geometric massive vector field $Q_{\mu}$ \cite{Vitagliano2010,Demir2020}. This vector field, a {\it geometric Proca field} as we will call it henceforth, is defined as 
 $Q_\mu \equiv \frac{1}{4} Q _{\mu \nu}^{~~~\nu}$ where 
 $Q_{\lambda\mu\nu}\equiv -{}^\Gamma \nabla_\lambda g_{\mu\nu}$ is the non-metricity tensor \cite{Demir2020, Buchdahl1979, Tucker1996,Obukhov1997,Vitagliano2010}. (We call non-metricity vector as geometric Proca to distinguish it from the generic Einstein-Proca system as well as the generic $Z^\prime$ bosons in the literature.) With a symmetric affine connection (torsion-free), one is left with a special case of  non-Riemannian geometries in which non-metricity $Q_{\lambda\mu\nu}$ is the only source of the deviations from the GR. This special case is the Weyl gravity \cite{Weyl0,Weyl1,Weyl2,Vitagliano2013} (see also gauge invariance analysis in \cite{gauge-th}). The geometric Proca is a direct signature of the Weyl gravity. More specifically, it is  signature of metric-incompatible symmetric connections (torsion-free).  It is not something put by hand. It is not something that comes from gauge theories either. It is a geometrical massive vector field that characterizes the Weyl nature of the geometry \cite{Weyl1,Weyl2}. It has been studied as vector dark matter in \cite{Demir2020}. Its couplings to fermions (quarks and leptons) were explored in \cite{dp-yeni} in regard to the black hole horizon in the presence of the Proca field \cite{Obukhov1997}.

One step further from the Palatini formalism with the geometric Proca $Q_\mu$ \cite{Demir2020,Vitagliano2010} is the inclusion of the metrical curvature $R_{\mu\nu}({}^g\Gamma)$ in addition to the already-present affine curvature 
${\mathbb{R}}_{\mu\nu}(\Gamma)$. Excepting the quadratic term ${\mathbb{R}}_{[\mu\nu]}(\Gamma) {\mathbb{R}}^{[\mu\nu]}(\Gamma)$ leading to the geometric Proca field, this combined metric-affine framework is the {\it metric-Palatini} gravity \cite{harko2012,Capozziello2013a,Capozziello2015}. 
This formalism, a direct combination of the metrical and Weyl gravities, helps relaxing the constraints on the Proca mass (as discussed in \cite{dp-yeni} in comparison to \cite{Demir2020}). The metric-Palatini gravity, excepting ${\mathbb{R}}_{[\mu\nu]}(\Gamma) {\mathbb{R}}^{[\mu\nu]}(\Gamma)$ term considered in the present work, has already been studied in regard to  dark matter \cite{Capozziello2012a}, wormholes \cite{Capozziello2012b}, and cosmology \cite{Capozziello2012c}. 

The gravity theory we explore in the present work is 
the metric-Palatini gravity extended with the \\ ${\mathbb{R}}_{[\mu\nu]}(\Gamma) {\mathbb{R}}^{[\mu\nu]}(\Gamma)$ invariant  and  a negative cosmological constant (CC). Indeed, as it was made clear in \cite{dp-yeni}, in the presence of the geometric Proca $Q_{\mu}$, the CC is a necessity for having black hole solutions. Our framework, which we call {\it extended metric-Palatini gravity} (EMPG),  possesses these four basic properties:
\begin{eqnarray}
\label{our-model}
&& {\it (i)}\ \text{It is linear in affine curvature scalar}\ g^{\mu\nu}{\mathbb{R}}_{\mu\nu}(\Gamma) ,\nonumber\\
&& {\it (ii)}\ \text{It is torsion-free}\ (\Gamma^{\lambda}_{\mu\nu}=\Gamma^{\lambda}_{\nu\mu}),  \\
&& {\it (iii)}\ \text{It involves the invariant}\ {\mathbb{R}}_{[\mu\nu]}(\Gamma) {\mathbb{R}}^{[\mu\nu]}(\Gamma), \nonumber\\
&& {\it (iv)}\ \text{It contains a negative CC}. \nonumber
\end{eqnarray}
Given these properties, the EMPG action takes the schematic form (whose exact form will be discussed in Sec. 2)
\begin{eqnarray}
\label{EMPG}
S[g,\Gamma]=\!\!\int\!\! d^4x \sqrt{-g}\! &&\Bigg \{\!``g^{\mu\nu}{R}_{\mu\nu}\left({}^g\Gamma\right)" + ``g^{\mu\nu}{\mathbb{R}}_{\mu\nu}\left(\Gamma\right)"\nonumber\\
&&+ ``{\mathbb{R}}_{[\mu\nu]}(\Gamma) 
 {\mathbb{R}}^{[\mu\nu]}(\Gamma)" + ``{\rm CC}" \!\Bigg \}
\end{eqnarray} 
in which ${R}_{\mu\nu}\left({}^g\Gamma\right)$ and ${\mathbb{R}}_{\mu\nu}\left(\Gamma\right)$ are, respectively, the metrical and affine Ricci curvatures. The EMPG is an Einstein-geometric Proca-Anti de Sitter (AdS) system. We will constrain its parameters and dynamics by using the observational data on black holes. In fact, in the literature, Einstein-Proca system (not the geometric Proca in the present work) has been 
analyzed for finding Reissner-Nordstr\"{o}m type spherically-symmetric vacuum solutions  \cite{Tresguerres1995a, 
Tucker1995, Vlachynsky1996, Macias1999},  for determining the role of the Proca field \cite{Bekenstein1971, Bekenstein1972, Adler1978}, for obtaining spherically-symmetric static solutions \cite{Frolov1978, Gottlieb1984, Leaute1985}, and for revealing the structure of the horizon radius
 \cite{Ayon1999, Obukhov1999, Toussaint2000}. In view of this rich literature, construction and  analysis of the Einstein-geometric Proca-AdS black holes  in the present work can probe affine curvature, with various applications in other astrophysical and cosmological phenomena. In essence, in the present work,  we are probing, for the first time in the literature, non-metricity tensor via the Einstein-geometric Proca-AdS black holes.

One of the main features of the metric theories of gravity is that the electromagnetic wave propagation will be affected by the spacetime curvature. The light deflection near the compact object due to the strong gravity can be used to test the corresponding metric theory. The first observational test of the general relativity proposed by Einstein in 1915 was performed using the gravitational lensing effect observed during the solar eclipse in 1919 by Arthur Eddington~\cite{lensing-0}. Later the gravitational lensing and electromagnetic wave propagation have been intensively studied by numerous authors in Refs.~\cite{Refsdal1966,Sidney1964,Refsdal1964,16}. Particularly, one may distinguish the gravitational lensing in weak field~\cite{weak1,weak2} and strong field~\cite{Bozza:2002b,Bozza2003aaa,Bozza:2005a,Eiroa:2004a,6,Virbha:2002a,Virbha:2009b,Islam_2020} regimes, where the deflection of the light from the initial line is not large and has large values, respectively. 

One of the consequences of the light propagation in curved spacetime is the appearance of the shadow of the black hole (BH). If the BH is located between the light source and the observer, then due to the capturing the part of the light by the central object observer detect the black spot in the bright background on the celestial plane. This black spot is referred to as BH shadow and first theoretically predicted by Synge~\cite{Synge:1960b} and further developed by Luminet~\cite{11} and Bardeen~\cite{Bardeen1973}. Even before the first ever observation of the BH shadow in 2019 by the Event Horizon Telescope collaboration~\cite{Akiyama19L1,Akiyama19L6} various authors have studied the shadow of the black hole/compact objects within general relativity and modified/alternative theories of gravity~(see, e.g.~\cite{Falcke00a,Bambi09a,Amarilla10a,Hioki09a,Abdujabbarov13a,Amarilla12a,Amarilla13a,Abdujabbarov16a,Atamurotov:2013sca,Tsukamoto18a,Abdujabbarov:2015rqa,Pantig2022,2021EPJP..136..436W,2020PhRvD.102j4032G,2021PhRvD.103j4047K,2020EPJC...80.1195H,2000CQGra..17..123D,2015MNRAS.454.2423A,2014PhRvD..89l4004G,Hou_2018,Atamurotov:2013dpa,Perlick18a,Cunha20a,2021MNRAS.tmp.1223A,2021PhRvD.103h4005B,Atamurotov:2013dpa,Atamurotov:2013sca,Abdujabbarov13a,Atamurotov:2015nra,Atamurotov:2015xfa,Papnoi:2014aaa,Babar:2020txt,Cunha_2017,atamurotov2021shadow,Atamurotov22kimet,Ghasemi-Nodehi:2020oiz,Sarikulov2F,Atamurotov22a,Mustafa2022shadow1}) and also the authors studied the effect of plasma on the BH shadow in  \cite{Perlick15a,Perlick17a,48,Atamurotov21axion,2021EPJC...81..987F,Bad21a,Ibrar2022shadow}. Observation data on the shadow of supermassive BHs at the center of M87\cite{Akiyama19L1} and Sgr A*\cite{Akiyama2022sgr} are used to get estimations and/or  constraints of BH parameters within different gravity models~\cite{2022arXiv220507787V}.  

In fact, despite large uncertainties of black hole shadow size including the mass and distance measurements, astrophysical observation related to black holes shadows in strong and weak gravity regimes is helpful in testing gravity theories. The first black hole shadow has been observed by Event Horizon Telescope observing at a wavelength of 1.3 mm, in 2019, the image of M87* (which has  6.5$\pm$0.7 billion solar masses, located at 16.8 kpc) with the angular size 42$\pm$3 $\mu$as with the observational resolution 20 $\mu$as \cite{Akiyama19L1}. Three years later, in 2022, the image of the central part of the Milky Way galaxy, known as Sgr A* (with $4^{+1.1}_{-0.6}$ million solar masses, located at about 8 kpc), has also been observed with shadow size 48.7$\pm$7 $\mu$as together with the radiation ring with the size 51.8$\pm$2.3 $\mu$as~\cite{Akiyama2022sgr}. The appearance of the radiation ring may be due to the presence of a plasma medium around Srg A*. Therefore, studies of the gravitational lensing effects of black holes including plasma is important. 

The present work  is organized as follows: In Sec. \ref{sec2}, we give a detailed analysis of the EMPG, starting with a ghost-free Lagrangian in the AdS background. We show that the EMGP gravity dynamically reduces to the GR plus a geometric Proca field  (the non-metricity vector from the rank-3 non-metricity tensor). In Sec. \ref{sec3}, we give static, spherically-symmetric solutions of the EMGP and apply our results to determine the Einstein-geometric Proca-AdS black hole solutions. In Sec. \ref{sec4}, we study photon motion around a compact object in the EMGP model to determine the optical properties of black holes and the black hole shadow. In Sec. \ref{sec5}, we study plasma effects on the black hole shadow. In Sec.~\ref{App}, we determine constraints on the EMPG parameters  using the observational image size of the supermassive black holes M87* and Sgr A*. Finally, we discuss our results in Sec. \ref{sec6}. (Throughout the work  we use a system of units in which $G=c=1$ and signature $(-,+,+,+)$) for the spacetime metric.)

\section{Einstein-Geometric Proca Model in A\lowercase{d}S background}\label{sec2}
The EMPG model, defined schematically in (\ref{EMPG}),  takes its exact form  \cite{dp-yeni,Demir2020} 
\begin{eqnarray}
\label{mag-action}
S[g,\Gamma]&&=\int d^4x \sqrt{-g} \Bigg \{
\frac{M^2}{2} {R}\left(g\right) + \frac{{\overline{M}}^2}{2} {\mathbb{R}}\left(g,\Gamma\right) \nonumber \\ 
&&+ \xi {\overline{\mathbb{R}}}_{\mu\nu}\left(\Gamma\right) {\overline{\mathbb{R}}}^{\mu\nu}\left(\Gamma\right) -V_0 + 
{\mathcal{L}}_{m}({}^g\Gamma,\psi) \Bigg \}, 
\end{eqnarray}
with an action based on the metric tensor $g_{\mu\nu}$ like the GR and on a symmetric affine connection $\Gamma^\lambda_{\mu\nu}=\Gamma^\lambda_{\nu\mu}$ unlike the GR. The affine connection is independent of the Levi-Civita connection
\begin{eqnarray}
{}^g\Gamma^\lambda_{\mu\nu} = \frac{1}{2} g^{\lambda \rho} \left( \partial_\mu g_{\nu\rho} + \partial_{\nu} g_{\rho\mu}-\partial_\rho g_{\mu\nu}\right)
\end{eqnarray}
generated by the metric $g_{\mu\nu}$. This connection sets the covariant derivative $\nabla_\mu$ such that $\nabla_\alpha g_{\mu\nu} = 0$. It also sets the Ricci curvature ${R}_{\mu\nu}\left({}^g\Gamma\right)$, which contracts to give the scalar curvature  $R(g) \equiv g^{\mu\nu}{R}_{\mu\nu}\left({}^g\Gamma\right)$.  The affine connection $\Gamma^\lambda_{\mu\nu}$, on the other hand, is independent of the metrical connection ${}^g\Gamma^\lambda_{\mu\nu}$, defines the covariant derivative   ${}^\Gamma \nabla_\mu$, and sets the Riemann curvature
\begin{eqnarray}
\label{affine-Riemann}
{\mathbb{R}}^\mu_{\alpha\nu\beta}\left(\Gamma\right) = \partial_\nu \Gamma^\mu_{\beta\alpha} - \partial_\beta \Gamma^\mu_{\nu\alpha} + \Gamma^\mu_{\nu\lambda} \Gamma^\lambda_{\beta\alpha} -\Gamma^\mu_{\beta\lambda} \Gamma^\lambda_{\nu\alpha}
\end{eqnarray}
with ${\mathbb{R}}^\mu_{\alpha\nu\beta}\left(\Gamma\right)=-{\mathbb{R}}^\mu_{\alpha\beta\nu}\left(\Gamma\right)$. Its contractions lead to two distinct Ricci curvatures  ${\mathbb{R}}_{\mu\nu}\left(\Gamma\right) = {\mathbb{R}}^\lambda_{\mu\lambda\nu}\left(\Gamma\right)$ and ${\overline{\mathbb{R}}}\left(\Gamma\right) = {\mathbb{R}}^\lambda_{\lambda\mu\nu}\left(\Gamma\right)$. The latter is actually the antisymmetric part of the former 
\begin{eqnarray}
{\overline{\mathbb{R}}}_{\mu\nu}\left(\Gamma\right) = {\mathbb{R}}_{[\mu\nu]} \left(\Gamma\right)  = \partial_\mu \Gamma^\rho_{\rho \nu} - \partial_\nu \Gamma^\rho_{\rho \mu}
\end{eqnarray}
and does necessarily vanish when the affine connection $\Gamma^\lambda_{\mu\nu}$ is replaced with the metrical one ${}^g\Gamma^\lambda_{\mu\nu}$. The total affine Ricci curvature leads to the affine scalar curvature $R(g,\Gamma)\equiv g^{\mu\nu}{\mathbb{R}}_{\mu\nu}\left(\Gamma\right)$ \cite{bauer-demir1,Demir2012}. 

The action (\ref{mag-action}) is composed of physically distinct parts \cite{dp-yeni}. The first part proportional to $M^2$ would be the usual Einstein-Hilbert action if $M$ were equal to the Planck mass $M_{Pl}$. The second term proportional to ${\overline{M}}^2$ is the standard Palatini action \cite{Palatini}, which leads to the  Einstein field equations with no need for extrinsic curvature \cite{york,gh}. The third term proportional to $\zeta$ was considered in both \cite{Vitagliano2010} and \cite{Demir2020}. The fourth term $V_0$ is the vacuum energy not considered in \cite{dp-yeni} (It is proportional to the CC in (\ref{EMPG}) in Introduction). The matter Lagrangian ${\mathcal{L}}_{m}({}^g\Gamma,\psi)$ governs the dynamics of the matter fields $\psi$ (matter sectors involving  $\Gamma^\lambda_{\mu\nu}$ (not  ${}^g\Gamma^\lambda_{\mu\nu}$) have been analyzed in \cite{Demir2020}). Our setup differs from the so-called metric-Palatini setup  \cite{harko2012,Capozziello2013a,Capozziello2015} by the third term proportional to $\zeta$ (and dropping of the higher powers of ${R}_{\mu\nu}\left({}^g\Gamma\right)$ and ${\mathbb{R}}_{\mu\nu}\left(\Gamma\right)$ in view of gravitational ghosts). 

In non-Riemannian geometries with symmetric connections ($\Gamma^\lambda_{\mu\nu}=\Gamma^\lambda_{\nu\mu}$) the torsion vanishes identically. Then, the difference between such geometries and the  metrical geometry of the GR \cite{Tucker1996,Obukhov1997,Vitagliano2010}
\begin{align}\Gamma^\lambda_{\mu\nu} - {}^g\Gamma^\lambda_{\mu\nu} =  \frac{1}{2} g^{\lambda \rho} ( Q_{\mu \nu \rho } + Q_{\nu \mu \rho } - Q_{\rho \mu \nu} )
\label{fark-connection}
\end{align}
is sourced by the non-metricity tensor
\begin{align}
Q_{\lambda \mu \nu} = - {}^\Gamma \nabla_{\lambda} g_{\mu \nu}
\label{non-met}
\end{align}
as a measure of the metric-incompatibility of the affine connection $\Gamma^\lambda_{\mu\nu}$. In fact, non-metricity is the main feature of the Weyl geometry \cite{Weyl1,Weyl2,gauge-th}. The use of the affine connection in (\ref{fark-connection}) in the action (\ref{mag-action}) and the use also of the equation of motion for the non-metricity tensor in (\ref{non-met}) leads one to the reduced action
\cite{Vitagliano2010,Demir2020,dp-yeni}
\begin{eqnarray}
S[g,Y,\psi] &=& \int d^4 x \sqrt{-g} \Bigg \{\frac{1}{16 \pi G_N} R(g)-V_0 \nonumber \\ &&- \frac{1}{4} Y_{\mu \nu} Y^{\mu \nu} - \frac{1}{2} M_Y^2 Y_{\mu} Y^{\mu} \nonumber \\
&&+ {\mathcal{L}}_{m} (g,{}^g \Gamma,\psi) \Bigg \}
\label{action-reduced}
\end{eqnarray}
in which $Q_\mu = Q _{\mu \nu}^{\nu}/4$ is the non-metricity vector,  $Y_\mu=2 \sqrt{\xi} Q_\mu$ is the canonical geometric vector field, $G_N=8\pi/(M^2 + \overline{M}^2)$ is Newton's gravitational constant, and
\begin{eqnarray}
M_Y^2 = \frac{3 \overline{M}^2 }{2 \xi}
\end{eqnarray}
is the squared-mass of the $Y_\mu$ (the  geometric Proca field).  We now bring the EMPG action (\ref{action-reduced}) into a more compact form
\begin{align}
S[g,Y] = \int d^4 x \sqrt{-g} \frac{1}{2 \kappa} \Bigg \{  R(g) - 2 \Lambda -  M_Y^2 \hat{Y}_{\mu} \hat{Y}^{\mu} \nonumber \\
- \frac{1}{2} \hat{Y}_{\mu \nu} \hat{Y}^{\mu \nu} \Bigg \}
\label{action-EGP}
\end{align}
in which $\kappa = 8 \pi G_N$, $\Lambda=\kappa V_0$ is the CC in (\ref{EMPG}), and $\hat{Y}_\mu \equiv \sqrt{\kappa} Y_{\mu}$ is the canonical dimensionless Proca field.

The actions (\ref{action-reduced}) and (\ref{action-EGP}) are readily recognized to belong to the Weyl geometry with Weyl vector $Y_\mu$ \cite{Weyl1,Weyl2,gauge-th}. It should nevertheless be kept in mind that neither the schematic action (\ref{EMPG}) nor the exact action (\ref{mag-action}) nor the reduced Weyl action (\ref{action-reduced})  are conformal-invariant. The reason is that the EMPG,  beginning from (\ref{EMPG}), is based on the Ricci curvatures and the CC not the Weyl tensor $W^\lambda_{\mu\nu\rho}$ whose quadratic is conformal-invariant. (Conformal transformations in Weyl gravity have been discussed in \cite{Demir2012,flanagan}.)

Variation of the action (\ref{action-EGP}) with respect to the metric $g_{\mu\nu}$ leads to the Einstein field equations
\begin{align}
R_{\mu \nu} - \Lambda g_{\mu \nu} - \hat{Y}_{\alpha\mu} \hat{Y}^{\alpha}_{\;\;\;\nu}  + \frac{1}{4} \hat{Y}_{\alpha \beta} \hat{Y}^{\alpha \beta} g_{\mu \nu} - M_Y^2 \hat{Y}_{\mu} \hat{Y}_{\nu} = 0,
\label{Einstein-eqns}
\end{align}
and its variation with respect to $\hat{Y}_\mu$  gives the Proca equation
\begin{align}
\nabla_\mu \hat{Y}^{\mu \nu} - M^2_Y \hat{Y}^\nu = 0,
\label{eom-Y}
\end{align}
describing $\hat{Y}_\mu$  as a free dimensionless massive vector field.

\section{Static Black Hole Solutions of the Einstein-Geometric Proca Model}\label{sec3}

In an attempt to find spherically-symmetric static solutions of the field equations in space coordinates $(r,\theta, \phi)$, we put forth the ansatz
\begin{align}
g_{\mu \nu} = \text{diag}(-h(r),\frac{1}{f(r)} ,r^2, r^2 \sin ^2 \theta).
\label{metric-sss}
\end{align}

Having done with the metric, the geometric Proca field $Y_\mu$ obeying the equation of motion (\ref{eom-Y}) can be taken  as a
purely time-like field 
\begin{align}
\hat{Y}_\mu = \hat{\phi}(r) \delta_\mu^0   , 
\label{proca-sss}
\end{align}
in agreement with  a spherically-symmetric background. With this time-like vector, the gravitational and geometric-Proca parts of the EMPG model are described by three real functions $h(r), f(r)$ and $\phi(r)$.

In order to have a dimensionless equation's system, we define the following dimensionless quantities:
\begin{align}
\hat{r} := \kappa^{-1/2} r,~ 
{\hat M}_Y^2 := \kappa M_Y^2~
\end{align}
So that along with ${\hat Y}_\mu$ the entire system gets expressed in terms of the dimensionless quantities.

The subtraction of the $(\mu \nu = 00)$ and $(\mu \nu = 11)$ components of the Einstein equations accordingly leads to the following equation
\begin{align}
\hat{\phi}^2 = \frac{1}{\hat{M}_Y^2 \hat{r}}(f h^\prime - f^\prime  h ).   
\end{align}

The $(\mu \nu = 22)$ component of the Einstein field equations reads as
\begin{align}
1- f -\frac{\hat{r}(h f)^\prime}{2 h} - \Lambda \hat{r}^2 -\frac{f \hat{r}^2}{2 h} \hat{\phi}^{\prime 2} = 0.
\end{align}
The Proca equation of motion (\ref{eom-Y}) becomes
\begin{align}
\frac{\sqrt{h f}}{\hat{r}^2} \Big ( \hat{r}^2 \sqrt{\frac{f}{h} } \hat{\phi}^\prime \Big )^\prime - \hat{M}_Y^2 \hat{\phi} = 0  \ .
\label{proca-eom}
\end{align}

In order to have an idea of what solution to expect we first solve the Proca equation (\ref{proca-eom}) in the flat spacetime limit ($f(\hat{r})=h(\hat{r})=1$). In this particular limit we find that
\begin{eqnarray}
\hat{\phi}^{(flat)}(\hat{r}) = c_1 \frac{e^{- {\hat{M}}_Y \hat{r}}}{\hat{r}} + c_2 \frac{e^{{\hat{M}}_Y \hat{r}}}{\hat{r}}\ ,
\label{Proca-flat}
\end{eqnarray}
which is what is expected of a Yukawa potential (massive vector field) in the flat spacetime.

Now, we turn to curved geometry in the weak field limit. In this limit, we take the metric functions in the form 
\begin{align}
h(\hat{r})=f(\hat{r}) = 1 + \frac{\hat{r}^2}{l^2}    , 
\end{align}
after letting $\Lambda = - \frac{3}{l^2}$ in (\ref{action-EGP})  in which $l$ stands for the AdS radius. In this case, the Proca equation (\ref{proca-eom}) takes the form 
\begin{align}
\frac{1}{\hat{r}^2} (1 + \frac{\hat{r}^2}{l^2})(\hat{r}^2 \phi')'= \hat{M}_Y^2 \hat{\phi} ,    
\end{align}
with the exact solution
\begin{align}
\hat{\phi}(\hat{r}) = \frac{l}{\hat{r}} c_1~ _2F_1 \left( \frac{-1 - \sigma}{4},\frac{-1 + \sigma}{4},\frac{1}{2},-\frac{\hat{r}^2}{l^2}\right) \nonumber \\ + c_2~ _2F_1 \left(\frac{1 - \sigma}{4},\frac{1 + \sigma}{4},\frac{3}{2},-\frac{\hat{r}^2}{l^2}\right) \ ,
\end{align}
where
\begin{align}\label{sigma}
\sigma=\sqrt{1 + 4 \hat{M}_Y^2 l^2} ~.    
\end{align}
The Breitenlohner-Freedman window of non-tachyonic negative mass-squared values lies in the following range
\begin{align}
m_{BF}^2 \leq \hat{M}_Y^2 < 0   ,
\end{align}
where
\begin{align}
m_{BF}^2 = - \frac{1}{4 l^2} 
\end{align}
is the Breitenlohner-Freedman mass bound which is required to avoid tachyonic run-away instabilities in the AdS background. The range of parameter $\sigma$ accordingly is
\begin{align}
0 \leq \sigma < 1 .  
\label{sigma-range}
\end{align}

Expanding $\hat{\phi}(\hat{r})$ around infinity the Proca field takes the following form
\begin{equation}\label{psi}
\hat{\phi}(\hat{r})=\frac{q_1}{\hat{r}^{\frac{1-\sigma}{2}}} + \frac{q_2}{\hat{r}^{\frac{1+\sigma}{2}}} ~.
\end{equation}
It is clear that in the Maxwell limit ($\sigma \rightarrow 1$) this Proca field behaves as $\hat{\phi}(\hat{r}) \rightarrow q_1 + \frac{q_2}{\hat{r}}$ and it means that $q_2$ has the meaning of an electromagnetic-like charge while $q_1$ represents a uniform potential.

Corresponding to the Proca field in (\ref{psi}), the metric potentials are expected to get modified as follows
\begin{align}
\label{metric-funcs-param}
f(\hat{r}) &= \hat{r}^2l^{-2} + 1 + \frac{n_1}{\hat{r}^{1 - \sigma}} + \frac{n_2}{\hat{r}}\ ,\nonumber\\
h(\hat{r}) &= \hat{r}^2l^{-2} + 1 +\frac{m_1}{\hat{r}^{1 - \sigma}}+\frac{m_2}{\hat{r}}\ ,
\end{align}
in which the higher inverse powers of $\hat{r}$ have been  ignored. En passant, one notices that the AdS geometry makes the Proca field to have the Coulombic form in (\ref{psi}) in place of the Yukawa form in (\ref{Proca-flat}) of the flat spacetime.  The parameters $n_1$, $n_2$ and  $m_1$ are found in terms of $q_1, q_2$ and $m_2$ by substituting (\ref{psi}) and (\ref{metric-funcs-param}) into the equations of motion 
\begin{align}
\label{param-solution}
&n_1 = \frac{(1 - \sigma)}{4} q_1^2\ , \nonumber \\
&n_2 = m_2 - \frac{(1 - \sigma)(1 + \sigma)}{6} q_1 q_2 \ , \nonumber\\
&m_1 = \frac{(1 - \sigma)}{3 - \sigma} q_1^2\ ,
\end{align}
where $q_1, q_2$ and $m_2$ are free parameters.

The ADM mass for this solution can be expressed as 
\begin{eqnarray}
\label{mass}
    M_{ADM}=\frac{1}{2}\! \left\{\!-m_2+q_1 q_2 \left[\gamma  \sigma +\frac{1}{3} (1-\sigma ) (\sigma +4)\right]\!\right\}
\end{eqnarray}
in which $\gamma$ is the coefficient of the surface term for the geometric Proca. Now, for simplicity, we take
 \begin{equation}
m_2=q_1q_2 \left(\gamma  \sigma +\frac{1}{3} (1-\sigma ) (4 + \sigma )\right)-2
\label{param-m2}
\end{equation}
after normalizing $m_2$ and $q_{1,2}$ in terms of the ADM mass $M_{ADM}$ ($M_{ADM}=1$).

For small ${\hat r}$ namely when ${\hat r} \ll l$, the metric functions in (\ref{metric-funcs-param}) take the form 

 \begin{eqnarray}
\label{metric-funcs}
f(\hat{r}) &=&  1 + \frac{(1 - \sigma) q_1^2}{4\hat{r}^{1 - \sigma}} \nonumber\\&+& \frac{-12 + q_1 q_2 \left ( 7 +6(\gamma-1) \sigma- \sigma^2 \right )}{6\hat{r}}\nonumber\\
h(\hat{r}) &=&  1 +\frac{ (1-\sigma ) q_1^2} {(3-\sigma ) \hat{r}^{1-\sigma }}\nonumber\\ &+&\frac{q_1 q_2 \left(\gamma  \sigma +\frac{1}{3} (1-\sigma ) (4 + \sigma )\right)-2}{\hat{r}}
\end{eqnarray}
after using the relations  in (\ref{param-solution}) and (\ref{param-m2}). Here, one recalls that ${\hat r}= \kappa^{-1/2} r$ is the dimensionless radius. 

\begin{figure}[h]
 \begin{center}
   \includegraphics[scale=0.8]{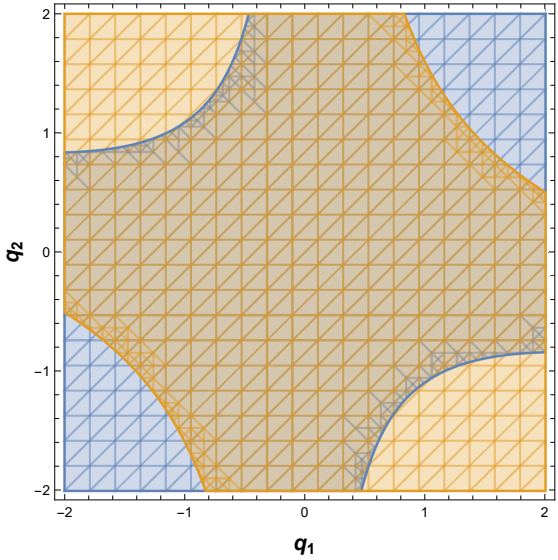}
  \end{center}
\caption{The allowed region for choosing the values of the parameters $q_{1}$ and $q_{2}$. The blue region is where $h(r)>0$ and the orange region is where $h(r)<1$. The values of $q_{1}$ and $q_{2}$ should be chosen from the intersection of these two regions. The values of $\gamma$ and $\sigma$ have been set to 1 and 0.8, respectively. }\label{plot:range}
\end{figure}

\begin{figure}[h]
 \centering
   \includegraphics[scale=0.65]{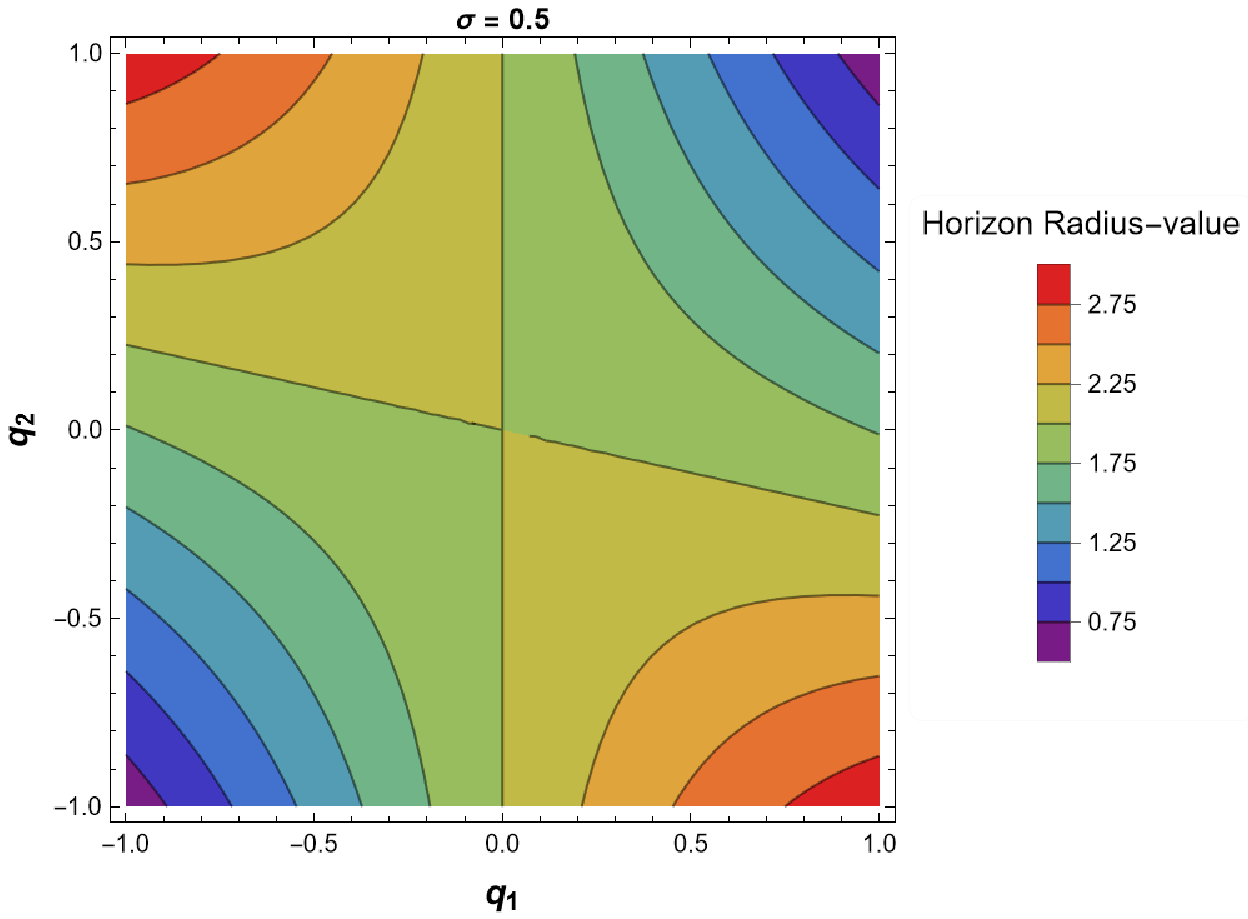}
   \includegraphics[scale=0.65]{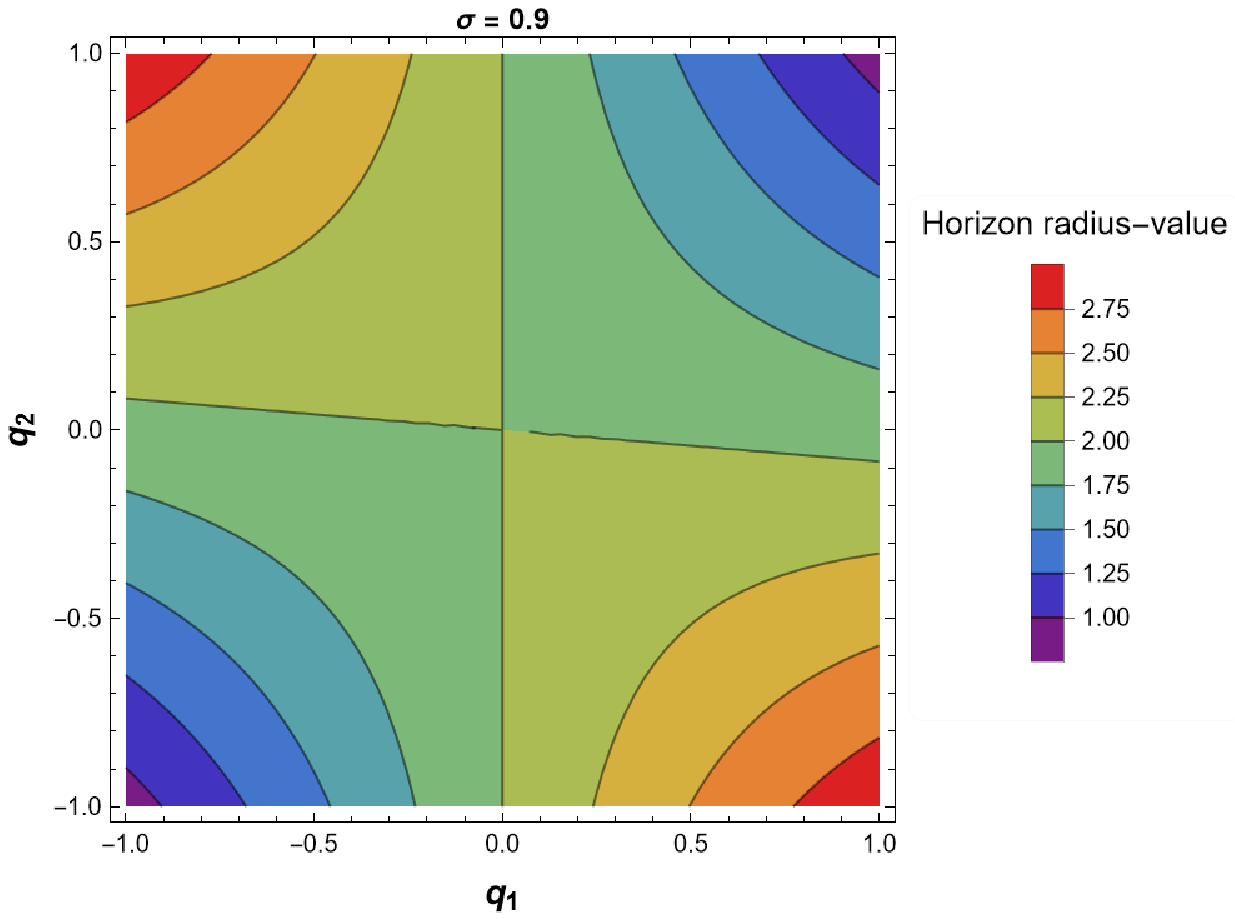}
 \caption{The dependence of horizon radius on $q_{1}$ and $q_{2}$ for two values of $\sigma=0.5$ and $\sigma=0.9$.}\label{plot:density}
\end{figure}

\begin{figure*}[ht!]
 \centering
   \includegraphics[scale=0.8]{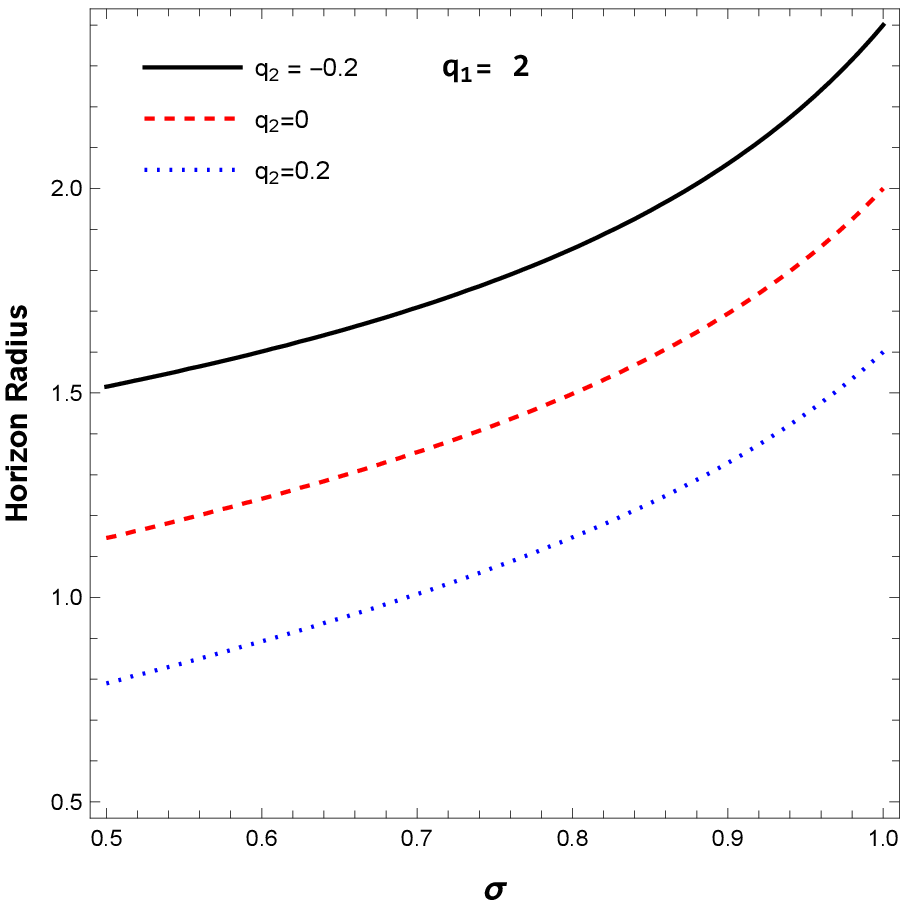}
   \includegraphics[scale=0.8]{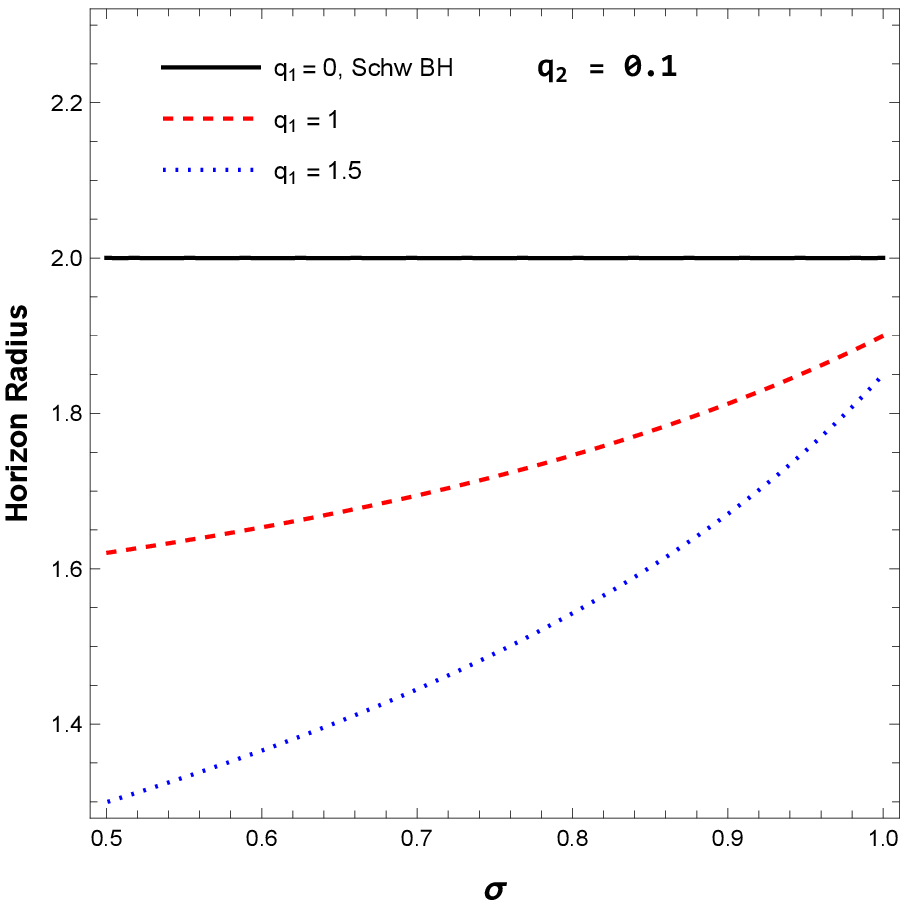}
\caption{Variation of horizon radius with increase of the parameter $\sigma$ for selected values of $q_{1}$ and $q_{2}$.  In the left plot $q_{1}$ is fixed and $q_{2}$ takes three different values while in the right plot, $q_{2}$ is fixed and $q_{1}$ takes three different values. If $q_1 \to 0$, the radius of horizon is the same as the horizon of Schwarzschild BH.}\label{plot:horizon2}
\end{figure*}

\begin{figure*}[ht!]
 \begin{center}
   \includegraphics[scale=0.8]{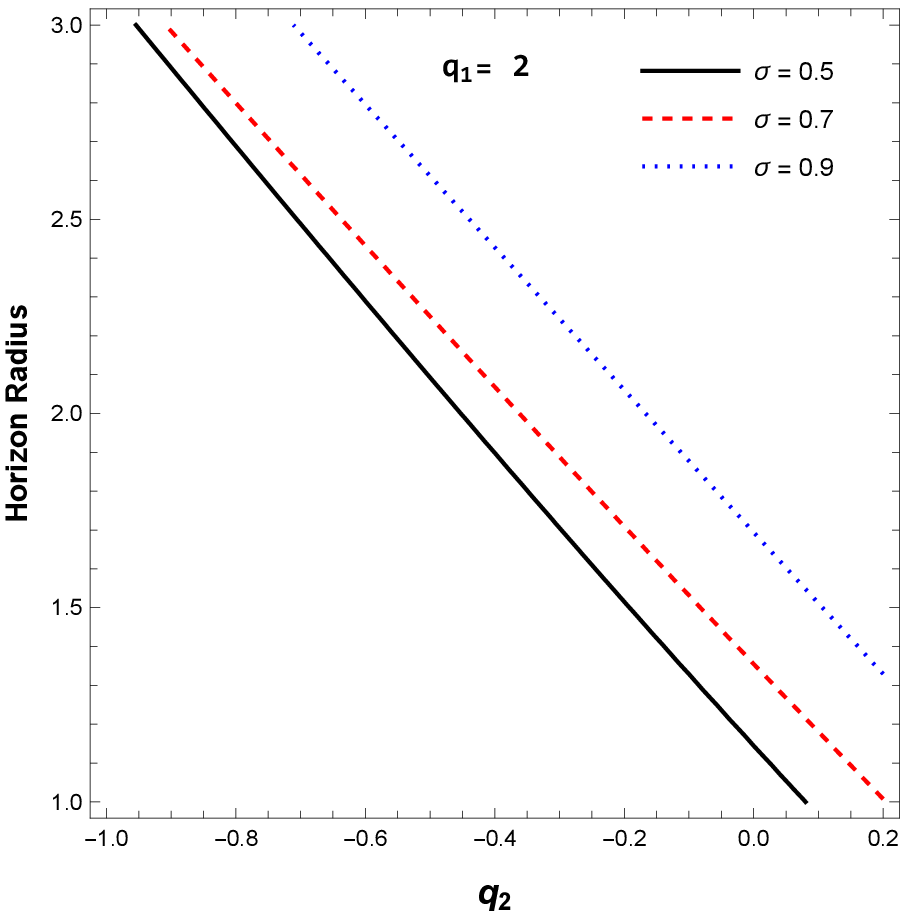}
   \includegraphics[scale=0.8]{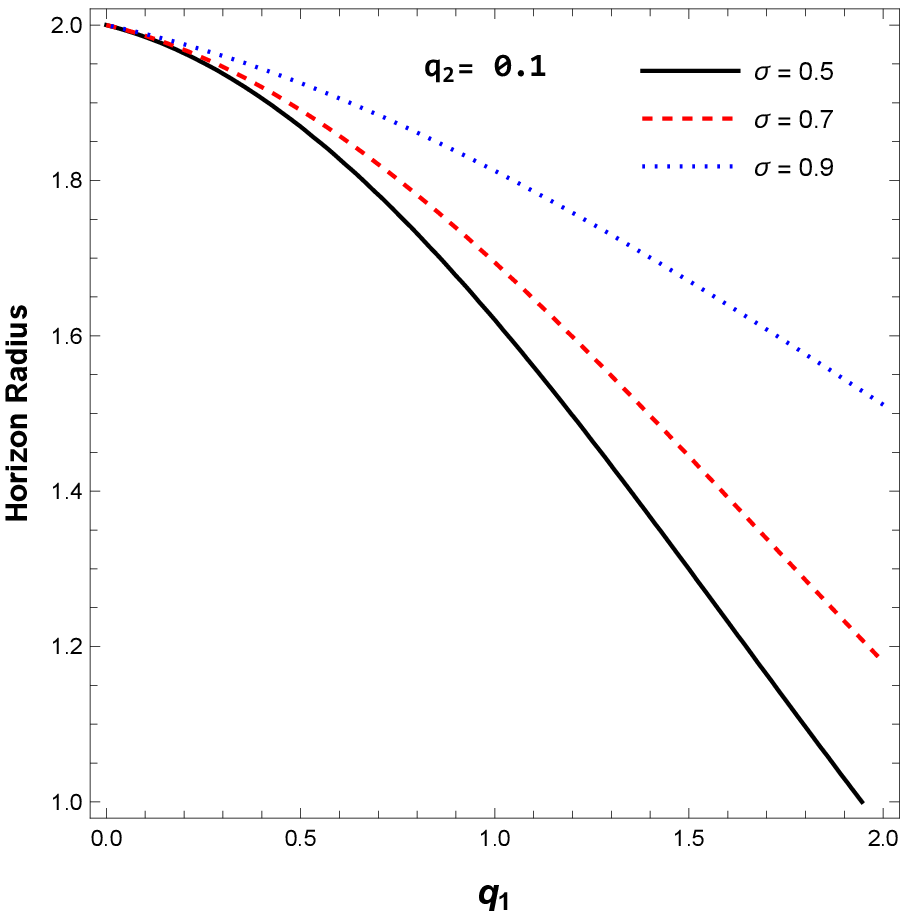}
\end{center}
\caption{The dependence of horizon radius on $q_{1}$ and $q_{2}$ for three values of $\sigma$. The solid black lines, dashed red lines and dotted green lines correspond to  $\sigma=0.5$, $\sigma=0.7$ and $\sigma=0.9$, respectively. In the left plot $q_{1}$ is fixed and in the right plot $q_{2}$ is fixed. }\label{plot:horizon3}
\end{figure*}

In order to study the horizon structure, first we need to find the allowed range of parameters. (To simplify the notation, here on we drop hats on parameters namely we use  $r$, $M_Y$ and $\phi$ to mean ${\hat r}$, ${\hat M}_Y$ and ${\hat \phi}$. We also set $\gamma=1$.) We find it by setting the condition $0<h(r)<1$. Fig~\ref{plot:range} shows this region for $\sigma=0.8$. The blue region shows where $h(r)>0$ and on the blue borders $h(r)=0$. In the orange region  $h(r)<1$ and on its borders $h(r)=1$. The intersection of these two regions is where $0<h(r)<1$, and we chose the value of $q_{1}$ and $q_{2}$ from this region. Fig~\ref{plot:density} shows how the horizon radius varies according to the parameters $q_{1}$ and $q_{2}$. When $q_{1}$ and $q_{2}$ both increase or decrease, the horizon radius decreases, but when one of $q_{1}$ or $q_{2}$ increases and the other one  decreases, the horizon radius increases. We see in both plots when $q_{1} = 0$, the horizon radius equals 2, which corresponds to the Schwarzschild case as we expected. It is also visible by setting $q_{1} = 0$ in equation (\ref{metric-funcs}). In Fig~\ref{plot:horizon2}, we can see how the horizon radius varies with respect to $\sigma$. In all cases, horizon radius increases as $\sigma$ increases except the case of $q_{1} = 0$, which represents the horizon radius for Schwarzschild black holes. Fig~\ref{plot:horizon3} shows that horizon radius decreases as $q_{1}$ or $q_{2}$ increases for selected values of $\sigma$. The lines intersect where $q_{1}=0$ in the right plot, but they descend and spread as $q_{1}$ grows.

The Kretschmann scalar (KS) $R_{\mu\nu\sigma\rho}R^{\mu\nu\sigma\rho}$ and the Ricci-squared (RS) scalar $R_{\mu\nu}R^{\mu\nu}$ possess lengthy expressions as invariant functions of ${r}$, $\sigma$, $q_1$ and $q_2$. It is clear from (\ref{metric-funcs}) that, for $q_1=0$,  the geometry reduces to Schwarzschild geometry namely $f({ r})=h({ r})=1-2/{ r}$ with a horizon at ${ r}=2$. In this sense, to reveal the geometry of the EMPG black hole near the Schwarzschild limit we expand the KS and RS scalars around $q_1=0$ up to quadratic order to find
\begin{align}
&R_{\mu\nu\sigma\rho}R^{\mu\nu\sigma\rho} = \frac{48}{r^6} \nonumber \\
&+ \frac{q_1}{36 r^6  (r-2)^4} \Bigg ( - \frac{1}{\sigma-3} \left [ 72 q_1 r^\sigma (r-2)^2 (\sigma -1)  \left ( 72   \right. \right. \nonumber \\
&- \left. \left. 98 \sigma + 22 \sigma^2 + r^2 (\sigma-3)(5 \sigma -7)   \right. \right.\nonumber \\
&+ \left. \left. r (\sigma (94 - 21 \sigma)\sigma - 77)  \right )\right ] + q_1 q_2^2 \left [ 192 (\sigma^2 -7)^2 \right. \nonumber \\
&+ \left. 6 r^4 (5 \sigma^4 - 46 \sigma^2 + 113) - 16 \hat{r}^3 (8 \sigma^4 - 103 \sigma^2 + 311) \right. \nonumber \\
&+\left. r^2 (241 \sigma^4 - 4130 \sigma^2 + 14257)  \right. \nonumber \\
&- \left. 16 r (19 \sigma^4 - 326 \sigma^2 + 1171)   \right ] + 48 q_2 (r-2)^2 \nonumber \\
& \times  \left [  r (9 r (\sigma^2 - 5) - 29 \sigma^2 + 173) + 24 (\sigma^2 - 7) \right ] \Bigg )
\label{Kretsch-scalar}
\end{align}
and 
\begin{align}
R_{\mu\nu}R^{\mu\nu} = \frac{q_1^2 q_2^2 (9 + r(3 r - 10))(\sigma^2 - 1)^2}{72 (r-2)^4 r^4}.   
\label{Ricci-scalar}
\end{align}
These invariants exemplify the KS and RS scalars near the physically-interesting domain of the Schwarzschild black hole. The KS scalar, as revealed by (\ref{Kretsch-scalar}), fully agrees with the Schwarzschild limit at the zeroth order in $q_1$. The RS scalar, on the other hand, vanishes up to the quadratic order 
$q_1$, in accordance with the Schwarzschild limit. In general, KS and RS behave differently in different geometries. In Schwarzschild solution, for instance, the spacetime is Ricci-flat and  KS remains as the indicator of the black hole singularity. In the EMPG this picture changes as a funtion of $q_1$ (and other parameters).

\begin{figure*}[h]
 \begin{center}
   \includegraphics[scale=0.6]{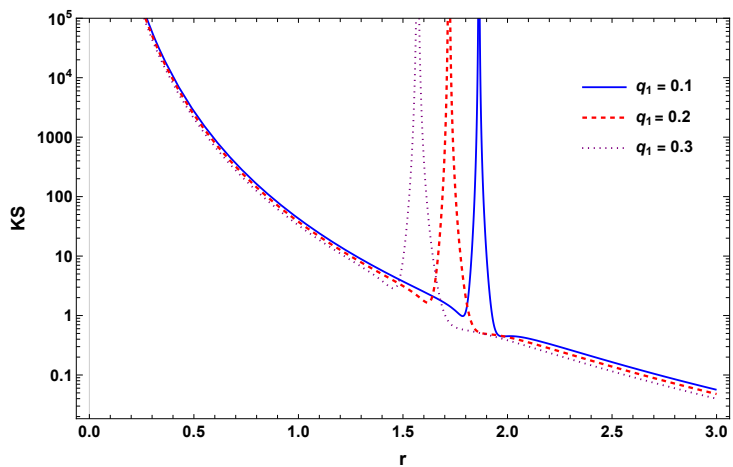}
  \end{center}
\caption{Variation of the Kretschmann scalar (KS) of the EMPG black hole with the radial distance $r$ for $q_2 =1$ and different values of  $q_1$ with $\sigma = 0$ and $\gamma=1$. The KS is seen to develop two singularities, one at the origin $r=0$ (independent of parameters) and the other at the horizon $r=r_H$ (dependent of the parameters). } \label{plot:KS}  
\end{figure*}

From hereon, we want to dwell on the singularity structure of the black hole by analyzing the KS in detail. The EMPG black hole develops two true singularities, one at the origin $r=0$ and the other at the horizon $r = r_H$. We call them primary and secondary singularities, respectively. (In the literature singularities at different $r$ values have already been discussed for the KS for the $5$-dimensional Schwarzschild-AdS spacetimes \cite{Seahra2005}, for the brane solutions of supergravity theory \cite{Peet2000} and for spherically-symmetric solutions of general relativity with scalar fields \cite{Stashko2021}.) Near the Schwarzschild limit $q_1=0$,  the KS expression in (\ref{Kretsch-scalar}) shows explicitly that the secondary singularity occurs at $r=2$, which is the event horizon of the Schwarzschild black hole.  The primary singularity remains just as in the Schwarzschild case. The secondary singularity, however, changes with changing parameters. Indeed, as we illustrate in Fig. \ref{plot:KS}, the secondary singularity varies with  different sets of $q_1$ and $q_2$. (Setting $\sigma = 0$ and $\gamma=1$).

It is worth nothing that the secondary singularity is a signature of the Proca field. Indeed, as $\sigma\rightarrow 1$ (for which $M_Y\rightarrow 0$) it is found that the KS reduces to  $\left({\rm KS} \right)_{EMPG} = \left({\rm KS} \right)_{Sch} + (12 q_1 q_2(q_1 q_2 -4))/r^6$ having a sole singularity at the origin just like Schwarzschild case  (with a $q_1$ and $q_2$ dependent residues, though). In this limit one recovers the electromagnetism-like geometric vector field. It is in this sense that the secondary singularity at the horizon $r=r_H$ is a direct signature of the Proca field. In fact, the plots in Fig. \ref{plot:KS} illustrate $\sigma=0$, a point far from the  electromagnetism-like limit of $\sigma=1$. Basically, sensitivity of the EMPG black hole on the parameters $\sigma$, $q_1$ and $q_2$ introduce a certain form of ``hair". It turns out the EMPG black hole possesses structures beyond the Schwarzschild case.


The stability of the EMPG black hole is another point to dwell on. In the literature, stability conditions for the spherically symmetric solutions of the extended gravity theories have been discussed in \cite{Sebastiani2013, Addazi2021}. The EMPG black hole admits the solution  
\begin{align}
 h(r) = f(r) = 1 - \frac{2 M_{BH}}{r} - \frac{\Lambda r^2}{3}
\end{align}
as follows from the metric potentials in (\ref{metric-funcs-param}) and (\ref{param-solution}) after ignoring the matter contribution by taking $\sigma = 1$ and after setting $m_2 = -2 M_{BH}$. As was shown in 
\cite{Sebastiani2013, Addazi2021}, near the extremal limit one is led to the Nariai black hole metric 
\begin{align}
d s^2 = e^{2 \rho} (- d \tau^2 + d x^2) + e^{-2 \phi} d \Omega^2  
\end{align}
in which 
\begin{align}
e^{2 \rho} = - \frac{1}{\Lambda \cos^2 \tau}, \; \;  e^{-2 \phi} = \frac{1}{\Lambda}
\end{align}
where $\tau = \arccos(\cosh{t})^{-1}$. Clearly, 
$- \infty < t < \infty$ corresponds to $- \pi/2 < \tau < \pi/2$ in the Nariai solution. It is necessary to study especially the perturbations $\delta\phi$ about the Nariai solution to determine stability of its spherically-symmetric structure:
\begin{align}
 \phi = \ln \sqrt{\Lambda} + \delta \phi
 \label{pertubations}
\end{align}
For $\delta \phi$, the perturbed Einstein field equations give the motion equation
\begin{align}
\frac{d^2 \delta \phi}{d t^2} + \tanh{t} \frac{d \delta \phi}{d t} - m^2 \delta \phi = 0
\label{eq-perturb}
\end{align}
in terms of the cosmological time $t$ and the mass parameter
\begin{align}
m^2 = \frac{2 (2 \alpha -1)}{3 \alpha}
\label{eq-perturb-mass}
\end{align}
where the parameter $\alpha$ is set by the curvature sector. It is clear that perturbations remain stable for $0 < \alpha < 1/2$ for which $2 \alpha -1 < 0$.  For $f(R)$ gravity is given by \cite{Sebastiani2013, Addazi2021}
\begin{align}
\alpha = \frac{\Lambda f_{R R}(R_0)}{f^\prime(R_0)}
\label{insta-param}
\end{align}
at constant curvature $R = R_0 = 4 \Lambda$. In this expression $f_R$ denotes derivative of $f$ with respect to $R$ and $f^\prime$ denotes derivative with respect to the radial coordinate $r$. At large times  $t \rightarrow \infty$  
$\tanh{t} \rightarrow 1$ and the perturbation in (\ref{eq-perturb}) behave as
\begin{align}
\delta \phi = \phi_0 ~ e^{\lambda_g t}
\label{perturb-sol1}
\end{align}
with the initial value $\phi_0$ and the Lyapunov exponent $\lambda_g = (-1 \pm \sqrt{1+4 m^2})/2$.

The EMPG model is an $f(R)=R-2 \Lambda$ theory. It is linear and it has therefore
\begin{align}
\alpha = 0   
\end{align}
as follows from (\ref{insta-param}). This $\alpha$ value is at the edge of the stability interval. It requires $m^2\rightarrow - \infty$ and this corresponds to a fully stable perturbation as follows from 
(\ref{eq-perturb}). The corresponding Lyapunov exponent implies periodic time-dependence for $\delta\phi$ and this dependence ensures stability. 
 
This stable solution of the perturbation shows that there arise no metric instabilities in the EMPG black hole (in the absence of matter).  The perturbation does not diverge. It is in this regard that the EMPG model is devoid of any metric instabilities. 

It is known that the critical impact parameter $b_C$ and the critical photon-trajectory are determined by the time-varying horizon radius which depends on the Lyapunov exponent as \cite{Sebastiani2013}
\begin{align}
b_C \simeq \sqrt{3}  r_{ph}  
= \sqrt{3}(3/2) r_{H}(t) = \sqrt{3} (3/2) r_{H}(0) e^{\lambda_g t} 
\label{impact-param}
\end{align}
so that for the EMPG model one expects no instability in the photon trajectory. Indeed, as expected from the imaginary value of $\lambda_g$ for $\alpha=0$, the photon radius shows small variations about the 
EMPG horizon radius $r_{H}(0)$.

The black holes can be destabilized also by non-minimal couplings of the Proca field \cite{Saenz2021}. This has been shown explicitly for the Schwarzschild solution in the presence of the non-minimally coupled Proca fields. But the EMPG model is a minimal Proca model (with no higher-order couplings to the self and to the curvature), and no instability is expected in this sense.

\section{Photon Motion around the BH}\label{sec4}

\subsection{Photon motion}

\begin{figure*}[ht!]
 \begin{center}
   \includegraphics[scale=0.8]{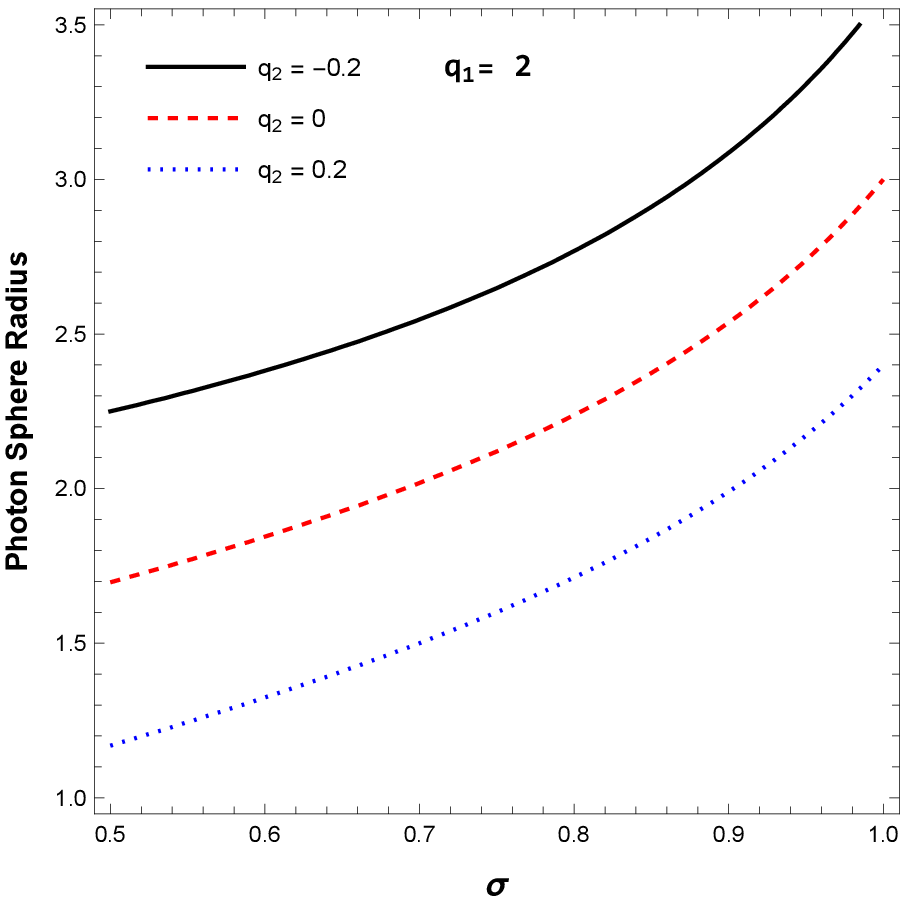}
   \includegraphics[scale=0.8]{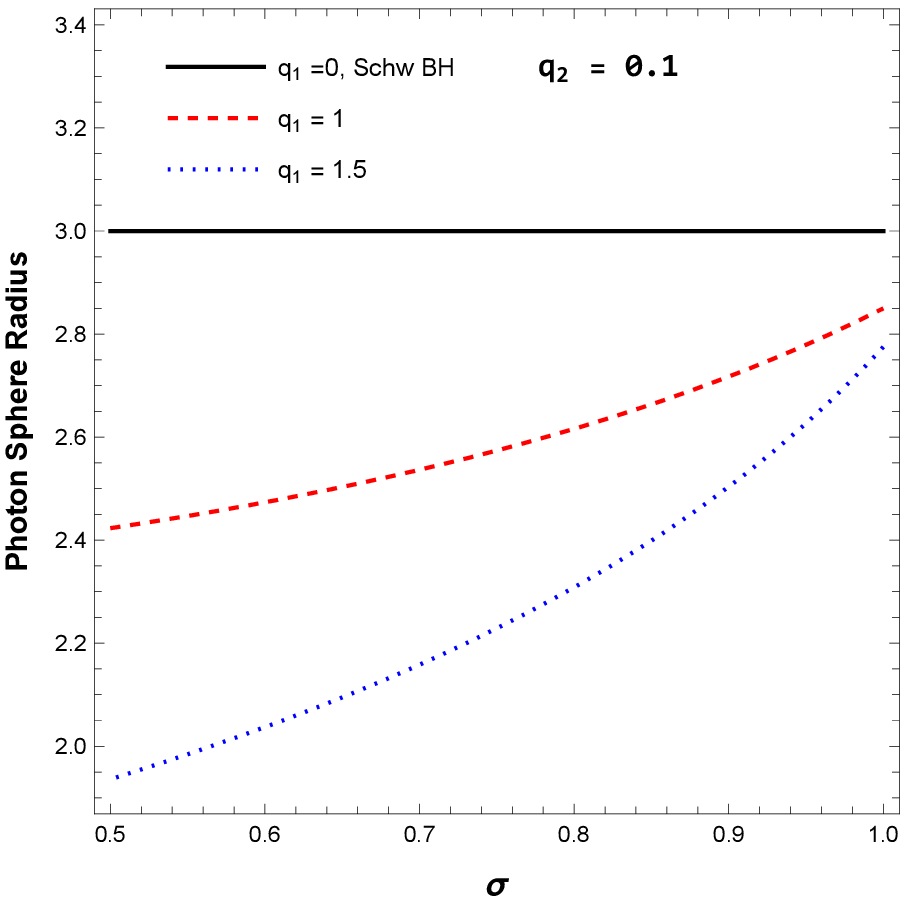}
  \end{center}
\caption{Variation of photon sphere radius with increase of the parameter $\sigma$, for selected values of $q_{1}$ and $q_{2}$.  In the left plot $q_{1}$ is fixed and $q_{2}$ takes three different values while in the right plot, $q_{2}$ is fixed and $q_{1}$ takes three different values. If $q_1 \to 0$, the radius of photon orbit is the same as with Schwarzschild BH's orbit.}\label{plot:photon1}
\end{figure*}

\begin{figure*}[ht!]
 \begin{center}
   \includegraphics[scale=0.8]{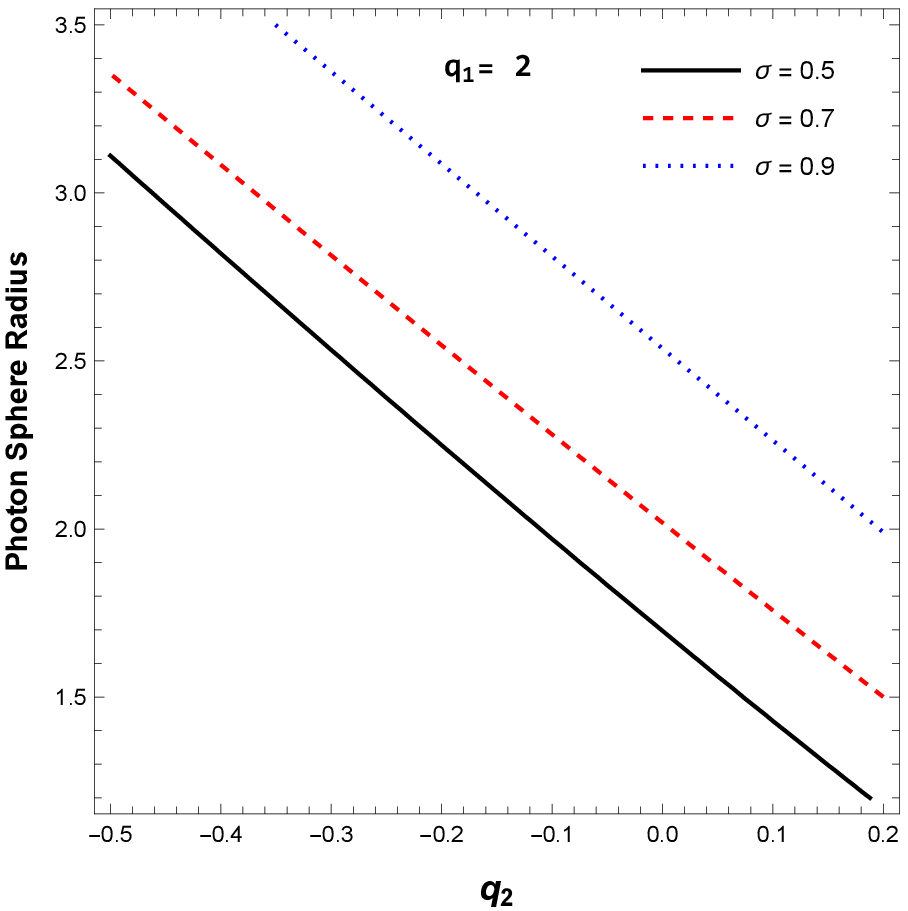}
   \includegraphics[scale=0.8]{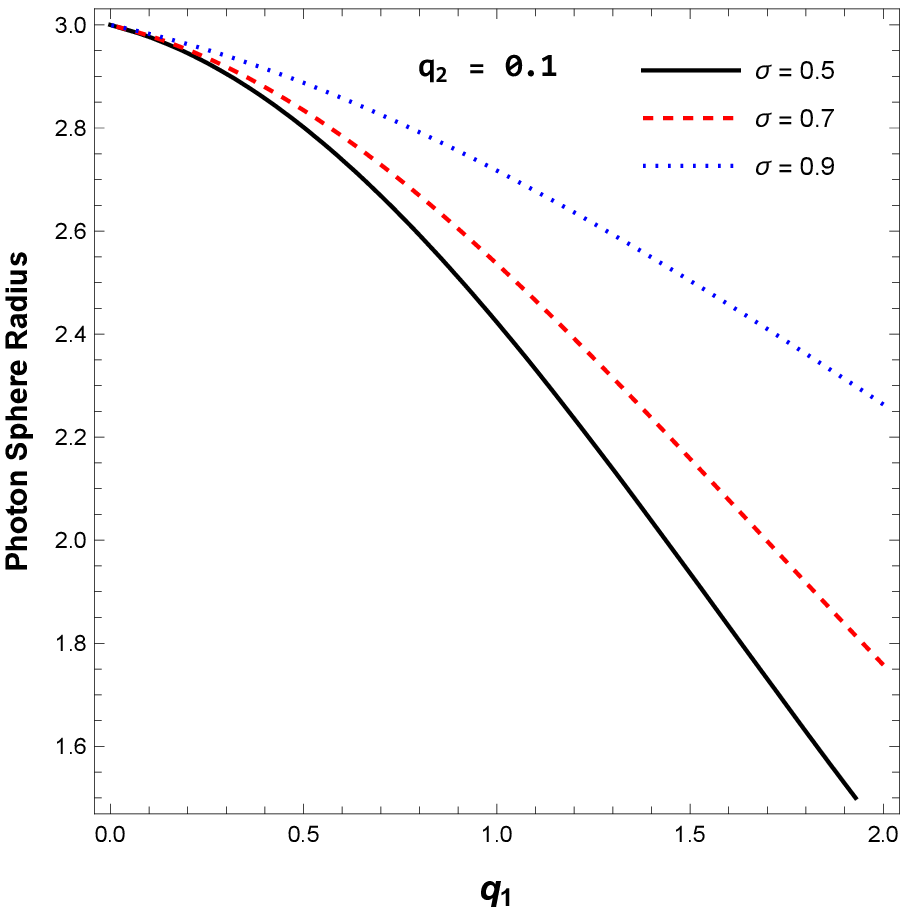}
  \end{center}
\caption{The dependence of photon sphere radius on $q_{1}$ and $q_{2}$ for three values of $\sigma$. The solid black lines, dashed red lines and dotted green lines correspond to  $\sigma=0.5$, $\sigma=0.7$ and $\sigma=0.9$, respectively. In the left plot $q_{1}$ is fixed and in the right plot $q_{2}$ is fixed.}\label{plot:photon2}
\end{figure*}

In this part we explore photon motion around BH in the EMPG model. We will use the Hamiltonian approach to investigate the photon motion. The Hamiltonian of the photons reads as 
\begin{equation}
\mathcal{H}=\frac{1}{2}\Big[g^{\alpha\beta}p_{\alpha}p_{\beta}\Big] , \label{eq:hamiltonnon}
\end{equation}
where $p^\alpha$ is the 4-momentum of the photons.
The components of the four velocity for the photons in the equatorial plane $(\theta=\pi/2,~p_\theta=0)$ are given by
\begin{eqnarray} 
\dot t\equiv\frac{dt}{d\lambda}&=& \frac{ {-p_t}}{h(r)}  , \label{eq:t} \\
\dot r\equiv\frac{dr}{d\lambda}&=&p_r f(r) , \label{eq:r} \\
\dot\phi\equiv\frac{d \phi}{d\lambda}&=& \frac{p_{\phi}}{r^2}, \label{eq:varphi}
\end{eqnarray}
where we used the relationship $\dot x^\alpha=\partial \mathcal{H}/\partial p_\alpha$. From Eqs. (\ref{eq:r}) and (\ref{eq:varphi}), we obtain a governing equation for the phase trajectory of light.
\begin{equation}
\frac{dr}{d\phi}=\frac{r^{2} p_r f(r)}{ p_{\phi}}.    \label{trajectory}
\end{equation}
Using the constraint $\mathcal H=0$, one can rewrite the above equation as~\cite{Perlick15a}
\begin{equation}\label{traj}
 \frac{dr}{d\phi}=\sqrt{r^2 f(r)}\sqrt{\gamma^2(r)\frac{p_t^2}{p_\phi^2}-1},
\end{equation}
where we defined
\begin{equation}
    \gamma^2(r)=\frac{r^2}{h(r)}.
\end{equation}
The radius of a circular orbit of light, particularly the one which forms the photon sphere of radius $r_{{ph}}$, is determined by solving the following equation ~\cite{Atamurotov22a}
\begin{equation}
\frac{d(\gamma^2(r))}{dr}\bigg|_{r=r_{{ph}}}=0. \label{eq:con}  
\end{equation}
The solution of this equation is depicted in Figs.~\ref{plot:photon1} and \ref{plot:photon2}. The plots for photon sphere radius exhibit exactly the same pattern as the horizon radius, as shown in Figs.~\ref{plot:horizon2} and~\ref{plot:horizon3}. {One may see that photon sphere enlarges as $\sigma$ increases except the case of $q_{1} = 0$, for which the photon sphere radius is $r_{ph}=3$ and represents Schwarzschild black holes.}

\subsection{Black hole shadow}

In this subsection we study the shadow of the BH described by the EMPG spacetime.
For the angular radius of the BH shadow we explore \cite{Perlick2022report,Atamurotov22a}
\begin{eqnarray}\label{shadow nonrotating1}
\sin^2 \alpha_{sh}=\frac{\gamma(r_{ph})^2}{\gamma(r_{obs})^2},
\end{eqnarray}
with
\begin{eqnarray}
\gamma(r)^2=\frac{g_{22}}{g_{00}}=\frac{r^2}{h(r)},\label{eq:h}
\end{eqnarray}
$\alpha_{sh}$ is the angular radius of the BH shadow, $r_{obs}$ is observer distance and $r_{ph}$ is already introduced in the previous subsection.  The Observer distance is very large but finite, and it has the value  $r_{obs}=D=8.3$ kpc for the Sgr A*\cite{Akiyama2022sgr} or $r_{obs}=D=16.8$ Mpc for the M87*\cite{Akiyama19L1}.

Now we combine Eqs.~(\ref{shadow nonrotating1}) and ~(\ref{eq:h}), and for an observer the Eq. (\ref{shadow nonrotating1}) takes the following form
\begin{eqnarray}\label{shadow nonrotating2}
\sin^2 \alpha_{sh}=\frac{r_{ph}^2}{h(r_{ph})}\frac{h(r_{obs})}{r^2_{obs}}.
\end{eqnarray}

One can find the radius of BH shadow for an observer at large distance using Eq. (\ref{shadow nonrotating2}) as \cite{Perlick2022report}

\begin{eqnarray}\label{shadow nonrotating3}
R_{sh}&\simeq&r_{obs} \sin \alpha_{sh} \simeq \frac{r_{ph}}{\sqrt{h(r_{ph})}}{\sqrt{h(r_{obs})}}.
\end{eqnarray}

\begin{figure*}[ht!]
 \begin{center}
   \includegraphics[scale=0.8]{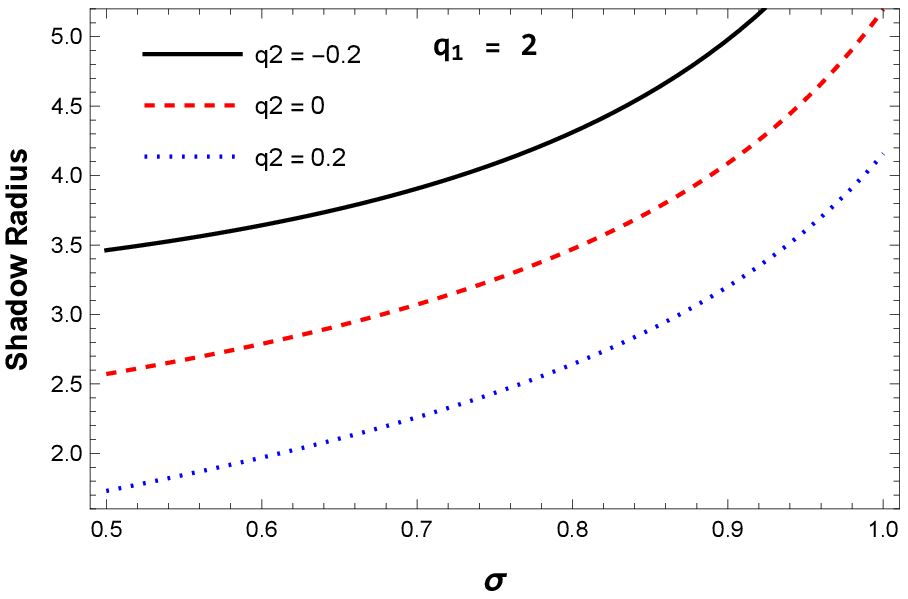}
   \includegraphics[scale=0.8]{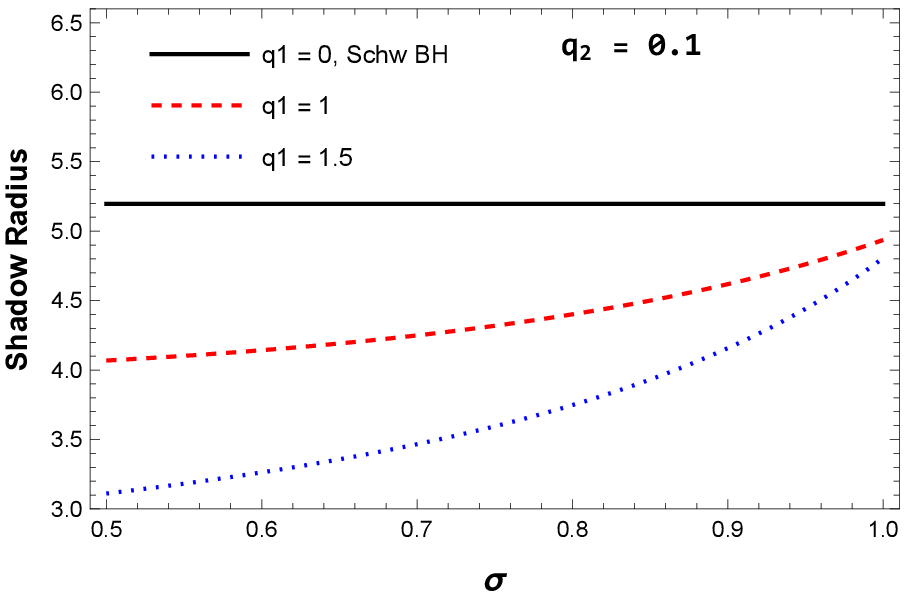}
  \end{center}
\caption{Variation of shadow radius with increase of the parameter
$\sigma$, for selected values of $q_{1}$ and $q_{2}$.  In the left plot $q_{1}$ is fixed and $q_{2}$ takes three different values while in the right plot, $q_{2}$ is fixed and $q_{1}$ takes three different values. If $q_1 \to 0$, the radius of BH shadow is the same as with Schwarzschild BH.}\label{plot:shadow}
\end{figure*}

\begin{figure*}[ht!]
 \begin{center}
   \includegraphics[scale=0.8]{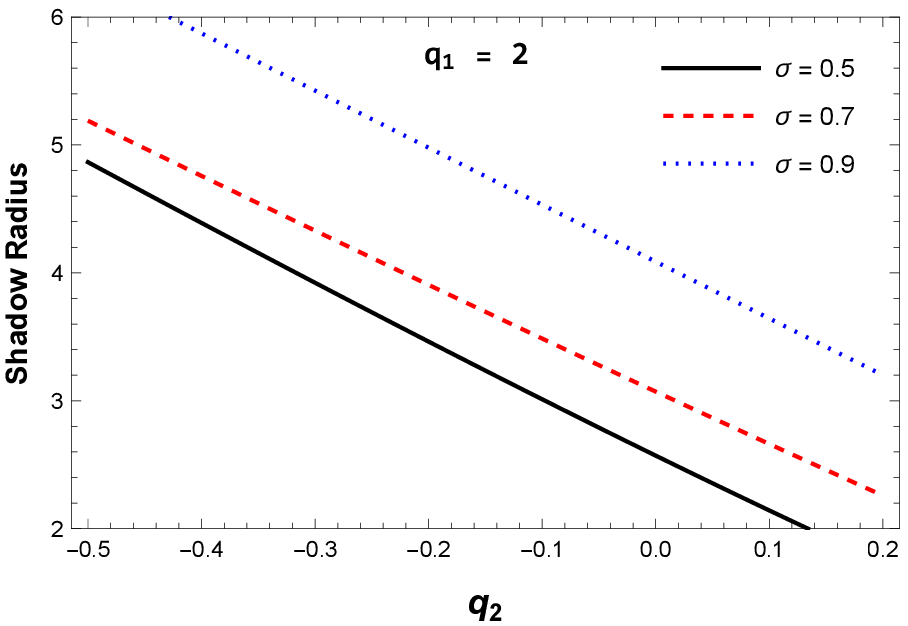}
   \includegraphics[scale=0.8]{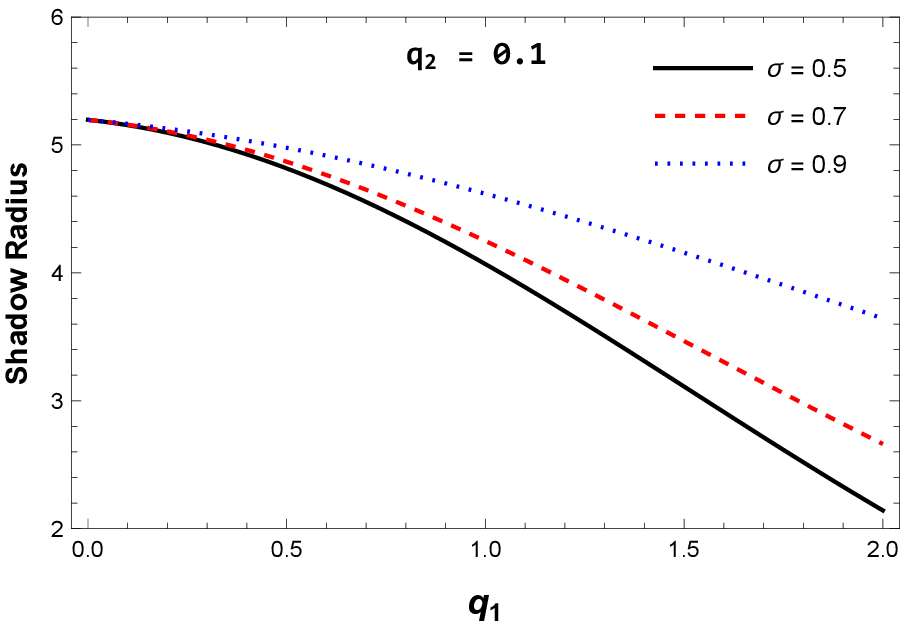}
  \end{center}
\caption{The dependence of shadow radius on $q_{1}$ and $q_{2}$ for three values of $\sigma$. The solid black lines, dashed red lines and dotted green lines correspond to  $\sigma=0.5$, $\sigma=0.7$ and $\sigma=0.9$, respectively. In the left plot $q_{1}$ is fixed and in the right plot $q_{2}$ is fixed. }\label{plot:shadow2}
\end{figure*}

Fig.~\ref{plot:shadow} shows how the shadow radius changes with respect to $\sigma$ for different values of $q_{1}$ and $q_{2}$. Increasing the value of $q_{1}$ or $q_{2}$, when one of them held fixed, decreases the shadow size. It is also worth mentioning that if we reverse the signs of $q_{1}$ and $q_{2}$ at the same time in Figs.~\ref{plot:horizon2}, \ref{plot:photon1} and \ref{plot:shadow}, we get exactly the same plots. This is because in the equation (\ref{metric-funcs}), in the lapse function, we have $q_{1}^2$ in the second term and the product $q_{1}q_{2}$ in the third term. Then, when $q_{1}$ and $q_{2}$ are both positive or negative, we get the same result. When they have opposite signs,  negative $q_{1}$ and positive $q_{2}$ give the same result as a positive $q_{1}$ and negative $q_{2}$ provided that $|q_1|$ and $|q_2|$ remain unchanged. Fig~\ref{plot:shadow2} shows the change of shadow size according to $q_{1}$ and $q_{2}$. Again, in the right plot, we see the shadow size of Schwarzschild black holes where the lines intersect at $q_{1}=0$.

\section{Photon Motion around the Black Hole In the Presence of Plasma}\label{sec5}

\subsection{Photon motion in the presence of plasma}
In this subsection we investigate the effect of a plasma environment on the photon motion around the BH. The Hamiltonian in the presence of a plasma environment is written as \cite{Perlick15a,Atamurotov21axion,Bad21a,Synge:1960b}
\begin{equation}
\mathcal{H}=\frac{1}{2}\Big[g^{\alpha\beta}p_{\alpha}p_{\beta}+\omega_p^2\Big], \label{eq:hamiltonian}
\end{equation}
where $\omega_p=\frac{4\pi e^2}{m_e}N_e(x)$  is the plasma frequency~\cite{Perlick15a}.
$N_e$ is the number density of electrons, $e$ and $m_e$ are the charge and mass of electron, respectively.

\begin{figure*}[ht!]
 \begin{center}
   \includegraphics[scale=0.8]{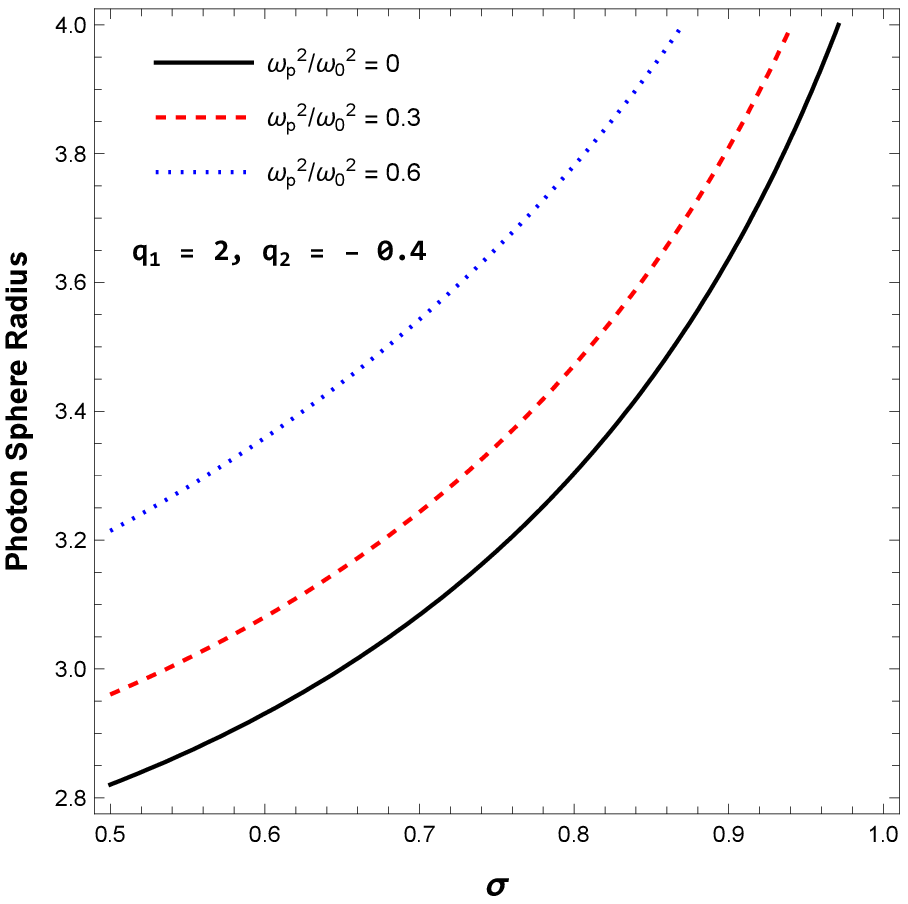}
   \includegraphics[scale=0.8]{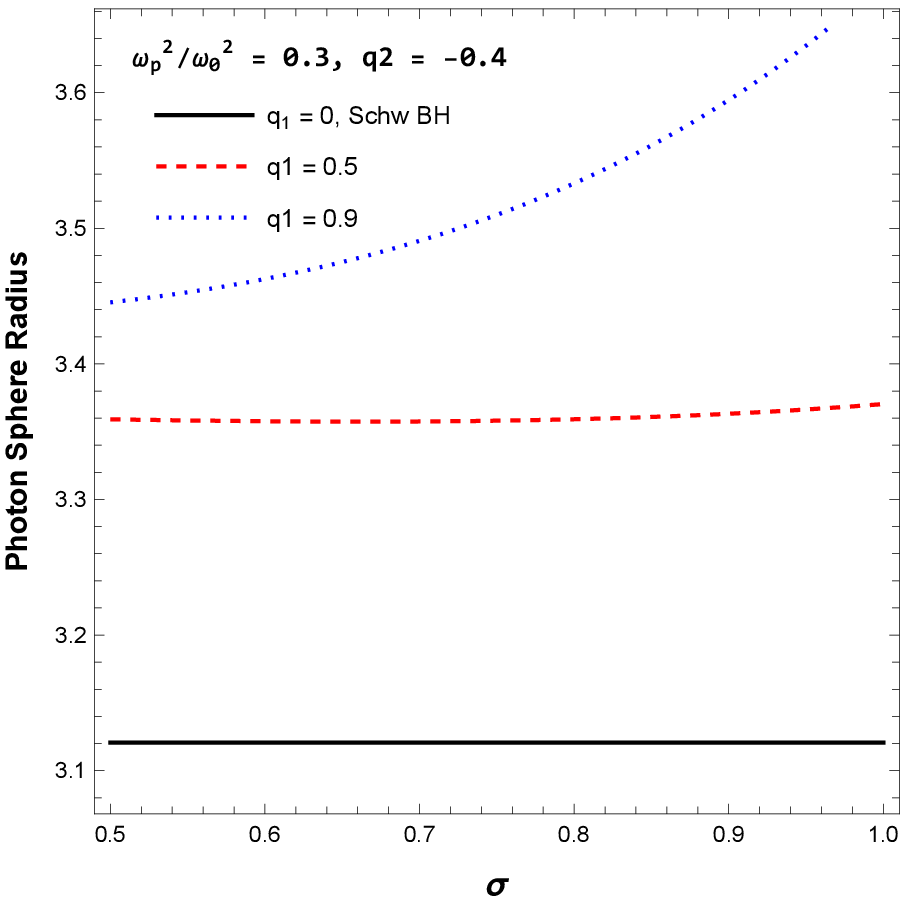}
    \includegraphics[scale=0.8]{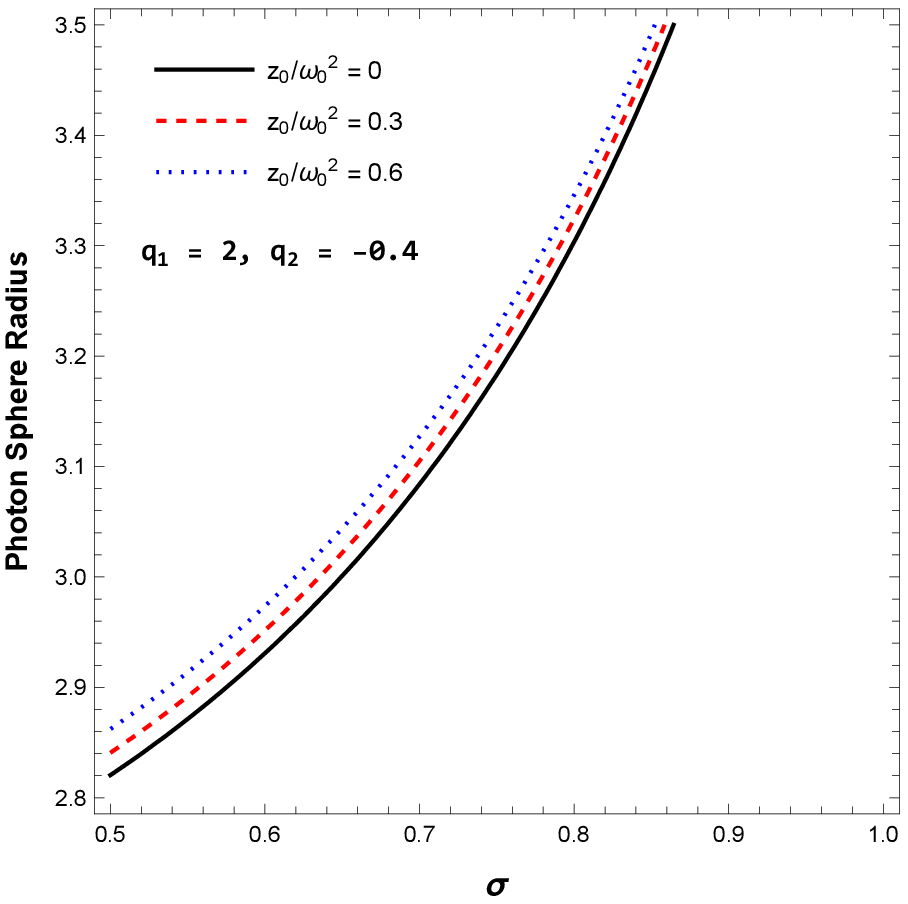}
   \includegraphics[scale=0.8]{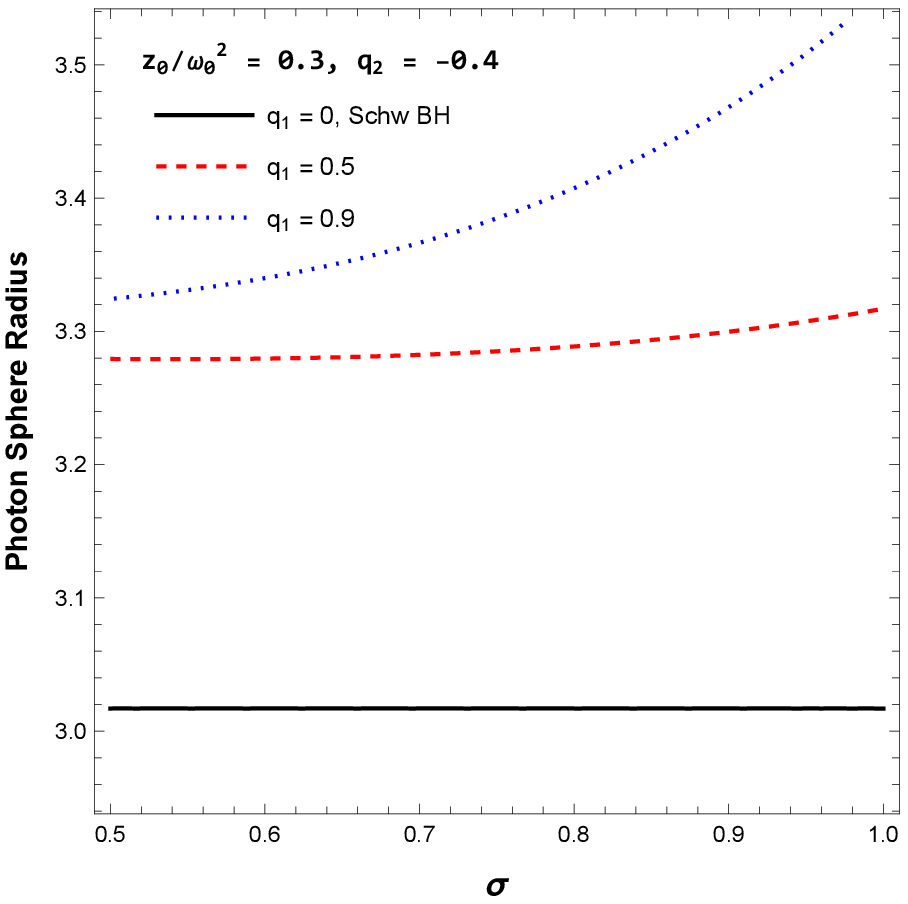}
  \end{center}
\caption{The dependence of photon sphere radius on $\sigma$ for selected values of  $q_{1}$ and $q_{2}$ in uniform (upper panels) and non-uniform (lower panels) plasma. In the left panels  $q_{1}$ and $q_{2}$ are fixed while plasma frequency is changing. In the right panels, plasma frequency and $q_{2}$ are fixed but $q_{1}$ is changing. If $q_1 \to 0$, the radius of photon orbit is the same as with Schwarzschild BH case.}\label{plot:plasma1}
\end{figure*}

The components of the four velocity for the photons and the governing equation for the phase trajectory of light are the same as equations (\ref{eq:t}) to (\ref{traj}) with the new $\gamma^2(r)$ modified as ~\cite{Perlick15a,Atamurotov21axion}
\begin{equation}
    \gamma^2(r)=r^2\left(\frac{1}{h(r)}-\frac{\omega_p^2}{\omega_0^2}\right), 
\end{equation}
where  $\omega_0=-p_t$. Then photon sphere is defined as a solution of the following equation
\begin{equation}
\frac{d}{dr}[\gamma^2(r)]\bigg|_{r=r_{{ph}}}=0. \label{eq:con}    
\end{equation}
Here we consider two cases: uniform plasma and non-uniform plasma. Uniform plasma  is defined as  ${\omega_p^2}/{\omega_0^2}= {\rm constant}$. For the non-uniform plasma, for the simplicity, we choose the following radial dependence of plasma frequency  \cite{Rog:2015a,Atamurotov21axion}:
\begin{equation}
    \omega_p^2(r)=\frac{z_0}{r}. 
\end{equation} 
$z_0$ is a free constant parameter~\cite{Rog:2015a,Atamurotov21axion}. Fig~\ref{plot:plasma1} shows the photon sphere size for uniform and non-uniform plasma. In both cases (left panels) photon sphere grows as ${\omega_p^2}/{\omega_0^2}$ or ${z_0}/{\omega_0^2}$ grow but the effect of ${\omega_p^2}/{\omega_0^2}$ on increasing the size of photon sphere in uniform case is stronger than the effect of ${z_0}/{\omega_0^2}$ in non-uniform case. 
The right panels show 
the effect of $q_1$ on the photon sphere size in the plasma environment. In contrast to the vacuum case, photon sphere enlarges as $q_1$ grows and becomes bigger than photon sphere of the Schwarzschild black holes. 

\subsection{Black hole shadow in plasma}

Now we investigate the radius of the shadow of the BH with the EMPG model in a plasma medium. The radius of the BH shadow in plasma can be 
written as~\cite{Perlick15a,Atamurotov21axion}

\begin{eqnarray}
R_{\text{sh}}&\simeq&\sqrt{r_{\text{ph}}^2\bigg[\frac{1}{h(r_{\text{ph}})}-\frac{\omega^2_p(r_{\text{ph}})}{\omega^2_0}\bigg]h(r_{obs})}.\label{eq.bhshadowplasma}
\end{eqnarray}
In the vacuum case $\omega_{\text{p}}(r)\equiv0$, we recover the radius of the shadow of the BH in the EMPG model without the plasma medium.

\begin{figure*}[th!]
 \begin{center}
   \includegraphics[scale=0.8]{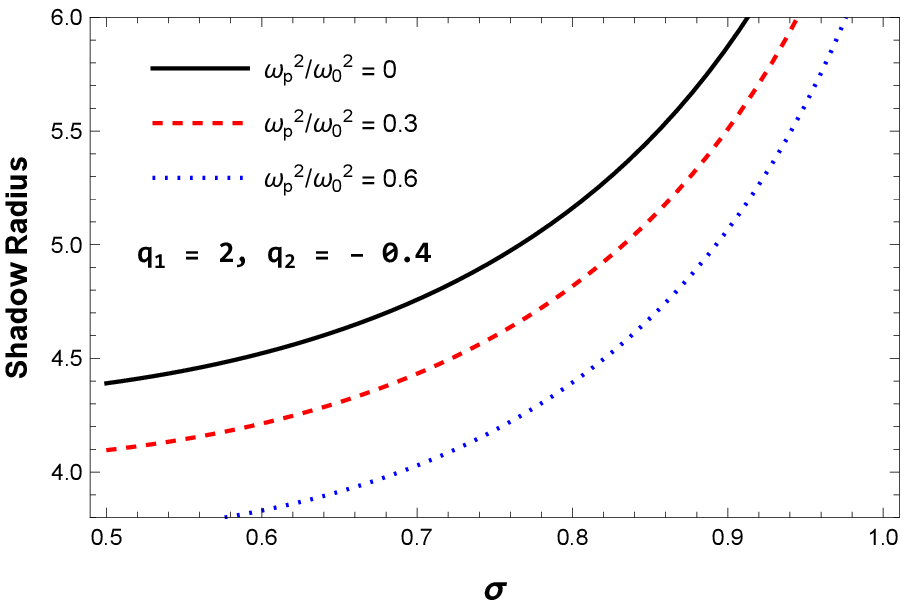}
   \includegraphics[scale=0.8]{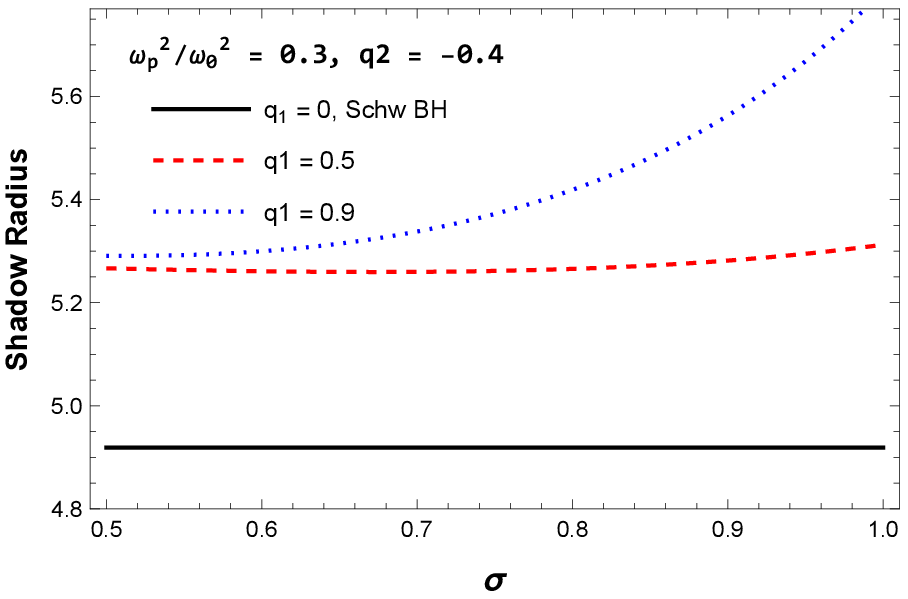}
   \includegraphics[scale=0.8]{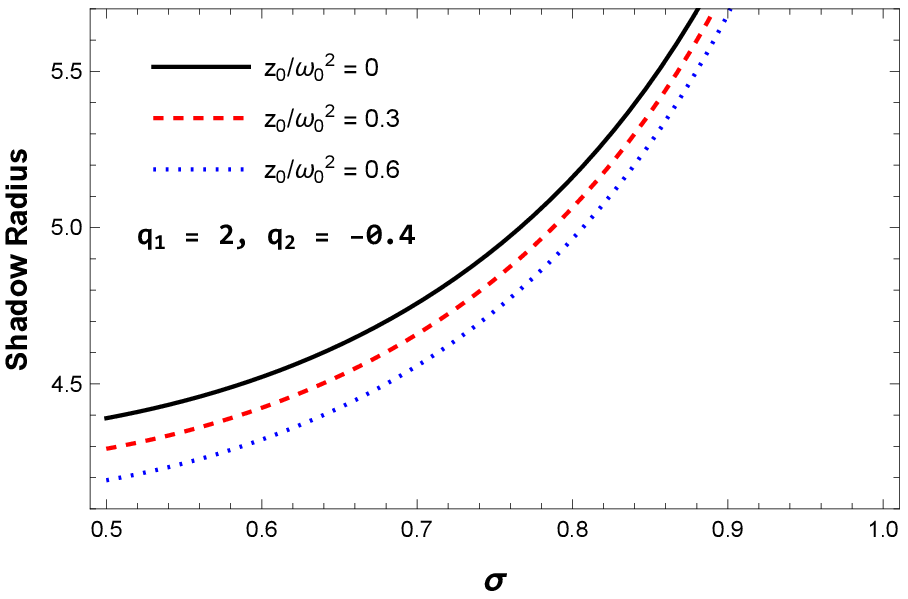}
   \includegraphics[scale=0.8]{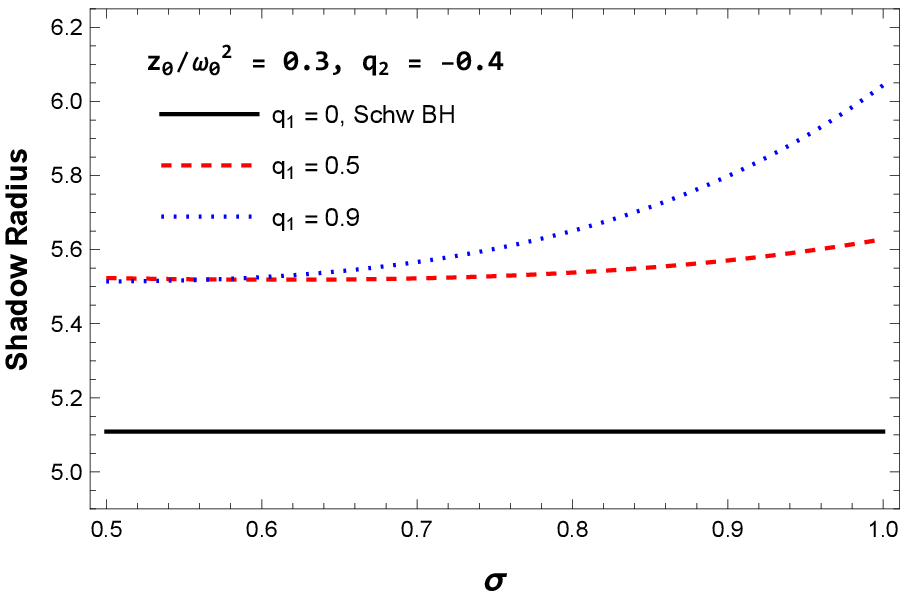}
  \end{center}
\caption{The dependence of shadow radius on $\sigma$ for selected values of  $q_{1}$ and $q_{2}$ in uniform (upper panels) and non-uniform (lower panels) plasma. In the left panels $q_{1}$ and $q_{2}$ are fixed while plasma frequency is changing. In the right panels, plasma frequency and $q_{2}$ are fixed but $q_{1}$ is changing.  If $q_1 \to 0$, the radius of BH shadow is the same as with Schwarzschild BH case.}\label{plot:plasma2}
\end{figure*}

Fig~\ref{plot:plasma2}  shows the shadow size for uniform and non-uniform plasma. In the left panels, we see the shadow size decreases as the plasma frequency increases, but we have seen that, for the same values of parameters, the photon sphere grows as plasma frequency grows in Figs~\ref{plot:plasma1}.  In other
words, plasma brings photon sphere and shadow close to
each other. Again one  can see the effect of $q_1$ on the shadow radius in the right panels and compare our result with the Schwarzschild case, which is shown in the figure coresponding to the case of $q_1=0$. It is also worth mentioning that for all the cases we studied till now, photon sphere and shadow had the same behaviour with respect to $\sigma$, for the same parameter choices.

\begin{figure*}[ht!]
 \begin{center}
   \includegraphics[scale=0.8]{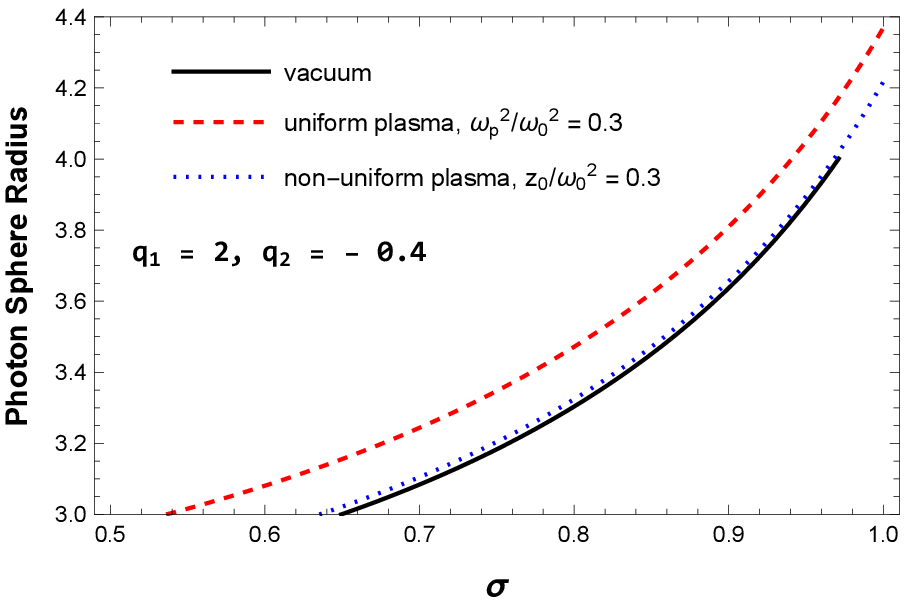}
   \includegraphics[scale=0.8]{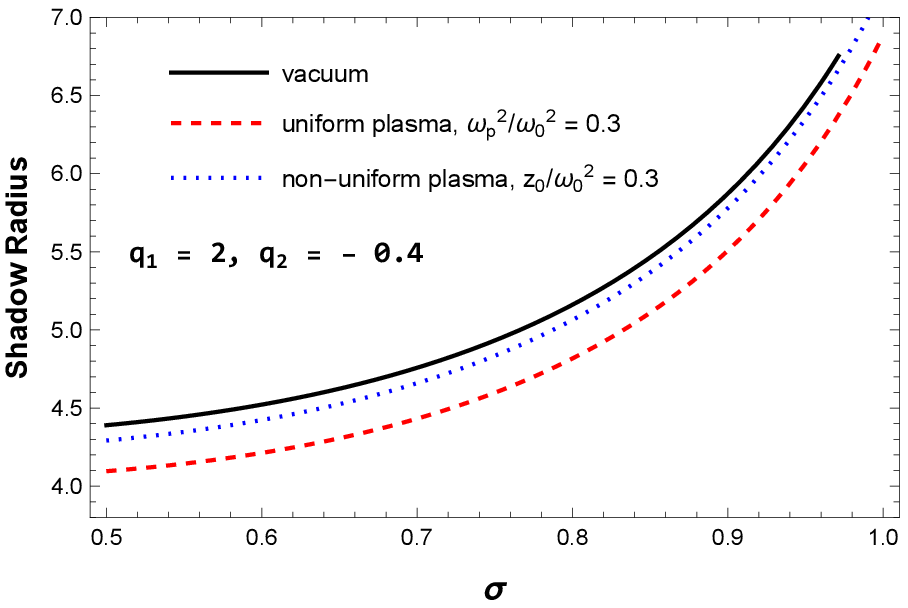}
  \end{center}
\caption{Comparing photon sphere and shadow radius  for vacuum, uniform plasma, and non-uniform plasma. The solid black lines, dashed red lines and dotted blue lines correspond to vacuum, uniform plasma, and non-uniform plasma cases, respectively.}\label{plot:all}
\end{figure*}

Fig.~\ref{plot:all} compares photon sphere and shadow for three cases: vacuum, uniform plasma, and non-uniform plasma. Here we can clearly see that plasma increases the photon sphere radius but decreases the shadow radius.

\section{Astrophysical Constraints from the EHT observations} \label{App}

In general, a BH shadow has two observable parameters: {\it (i)} shadow radius, and {\it (ii)} its distortion from the circle shape. Observations of the supermassive black holes (SMBHs) M87* and Sgr A* have revealed that their shadow shape are almost circular with a very small distortion parameter (with about $1-5\ \%$ variation). Also revealed is that shadows of both SMBHs have some common features like for example  shadow's center has some brightness depression and both are nearly-circular rings. As a result, the EHT collaboration has determined sizes of these BHs  by assuming ring-shaped structures. Here, in this section, we will explore potential constraints on the Proca field parameters using the Hioki\& Maeda method and the EHT observations of the shadows cast by SMBHs M87* and Sgr A*. Throughout our analysis will be based on the Einstein-geometric Proca BH solution.

\begin{figure*}[ht!]
 \begin{center}
   \includegraphics[scale=0.8]{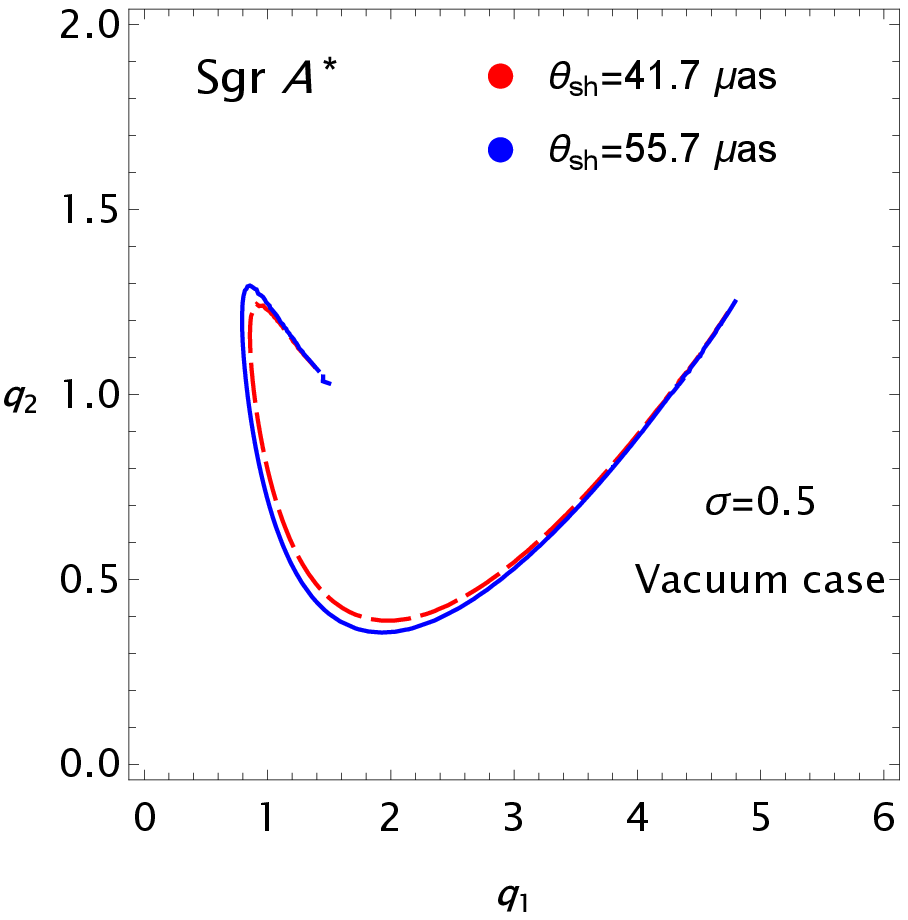}
   \includegraphics[scale=0.8]{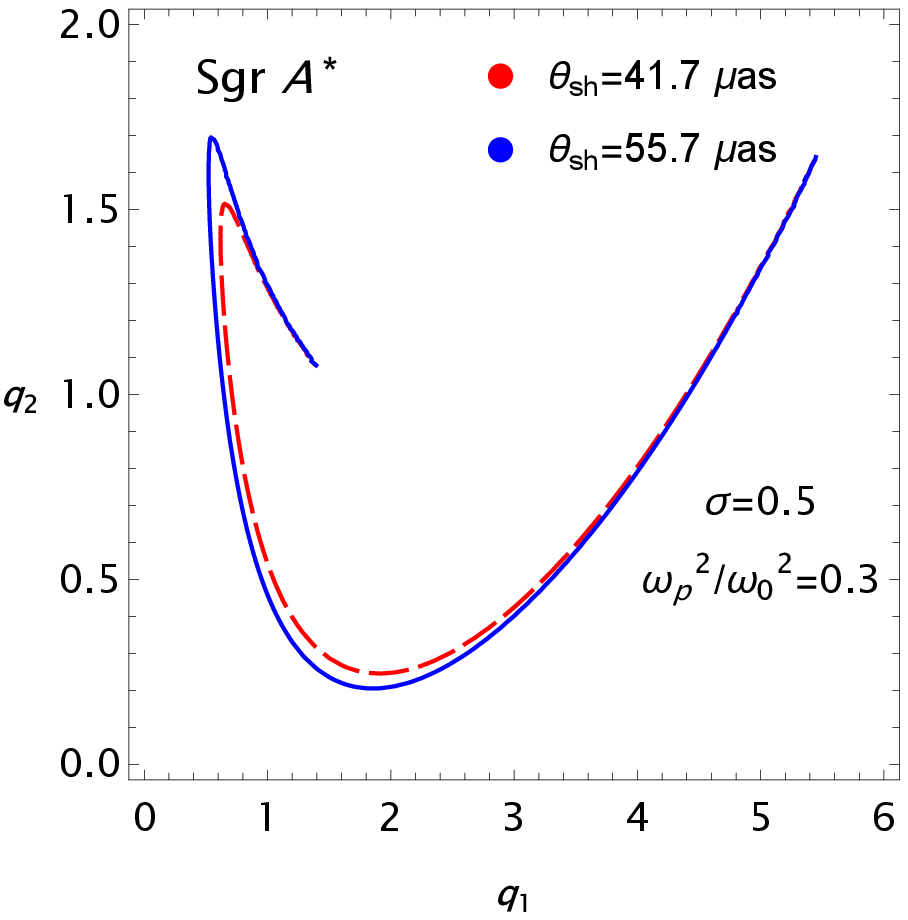}
   \includegraphics[scale=0.8]{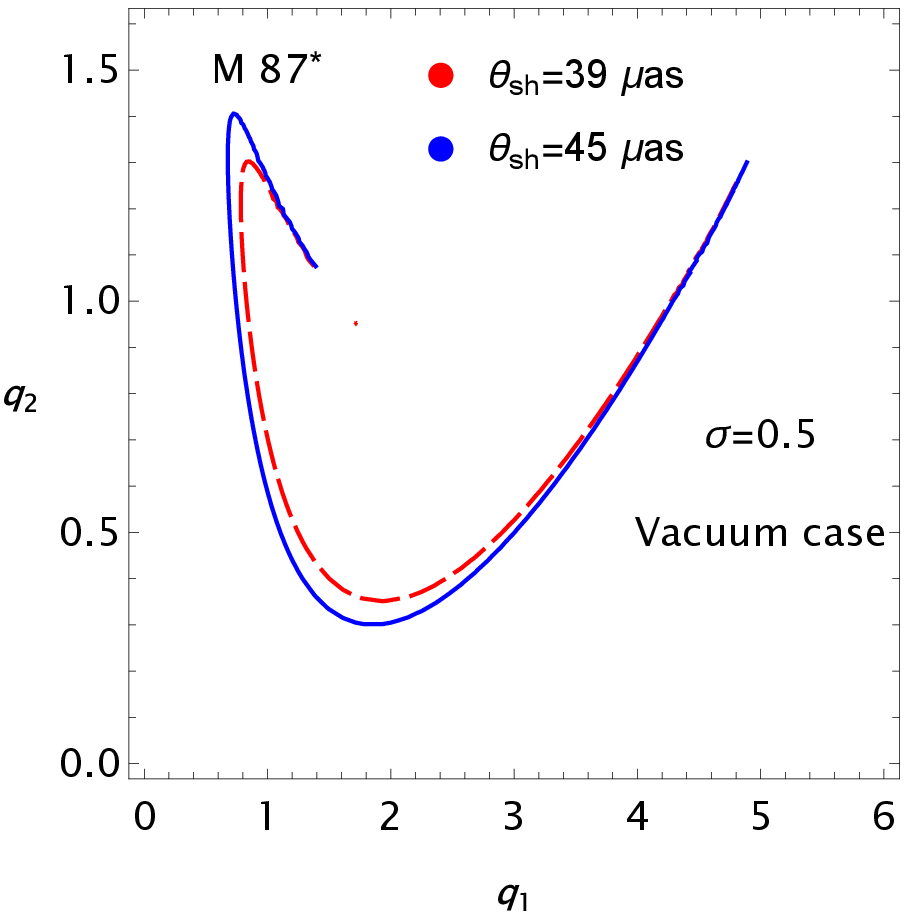}
   \includegraphics[scale=0.8]{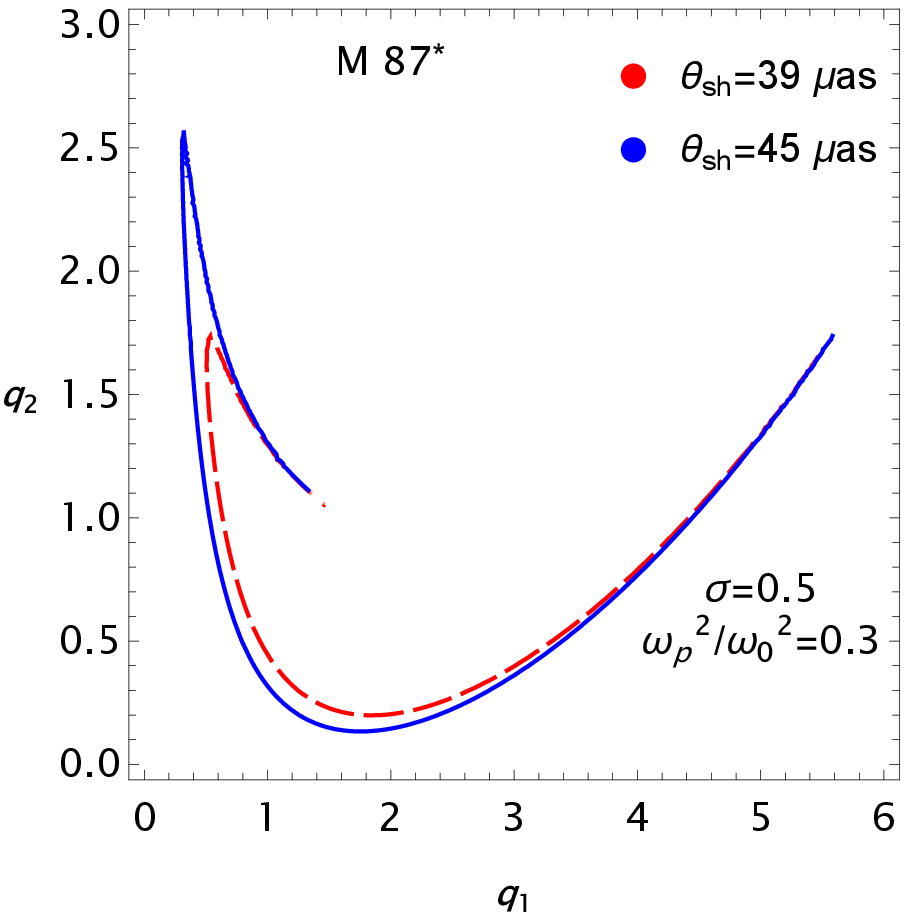}
  \end{center}
\caption{Constraints on possible values of the parameters $q_1$ and $q_2$ for $\sigma=0.5$. Here, we used the observational angular size of SMBHs M87* and SgrA* as $\theta_{sh}=42\pm3 \mu$as and $\theta_{sh}=48.7\pm 7 \mu$as, respectively. }\label{plot:M87SgrA}
\end{figure*}

\begin{figure*}[htt!] \begin{center}    \includegraphics[scale=0.8]{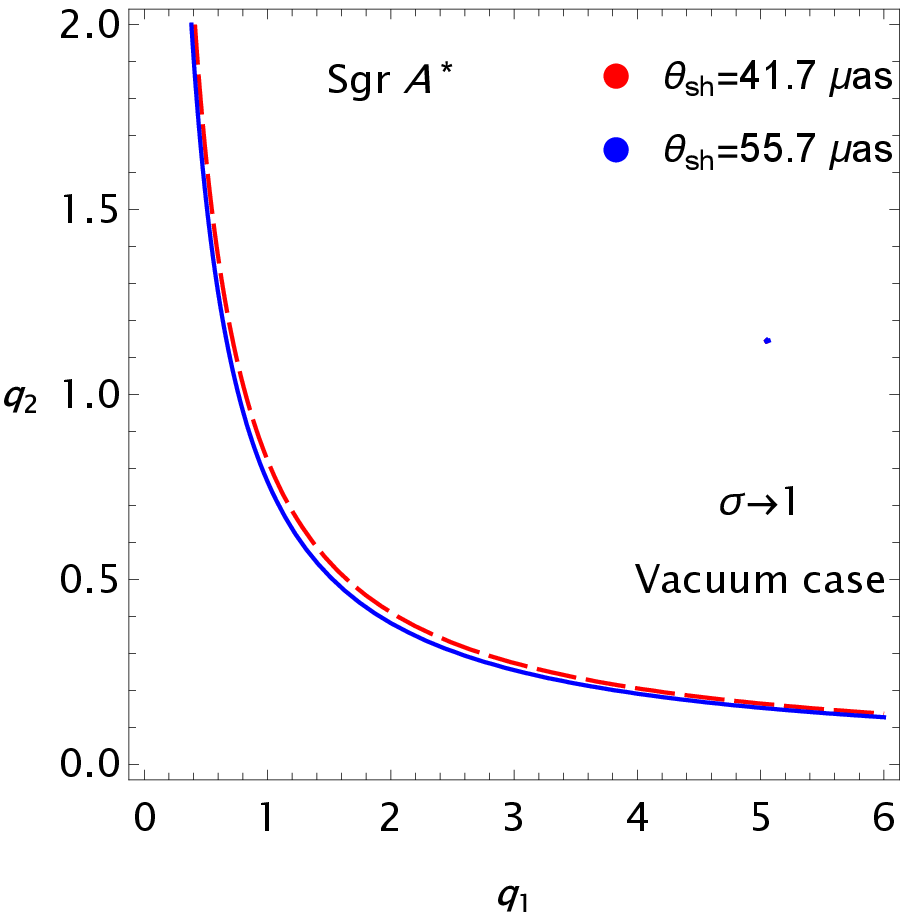}       \includegraphics[scale=0.8]{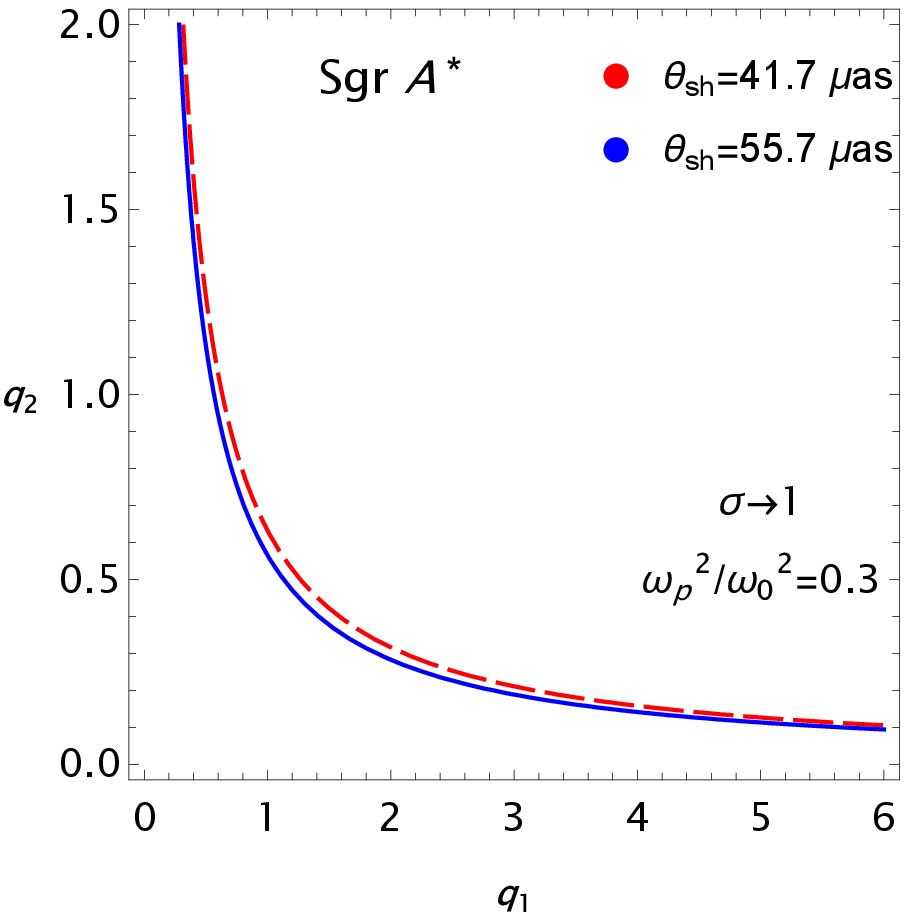}    \includegraphics[scale=0.8]{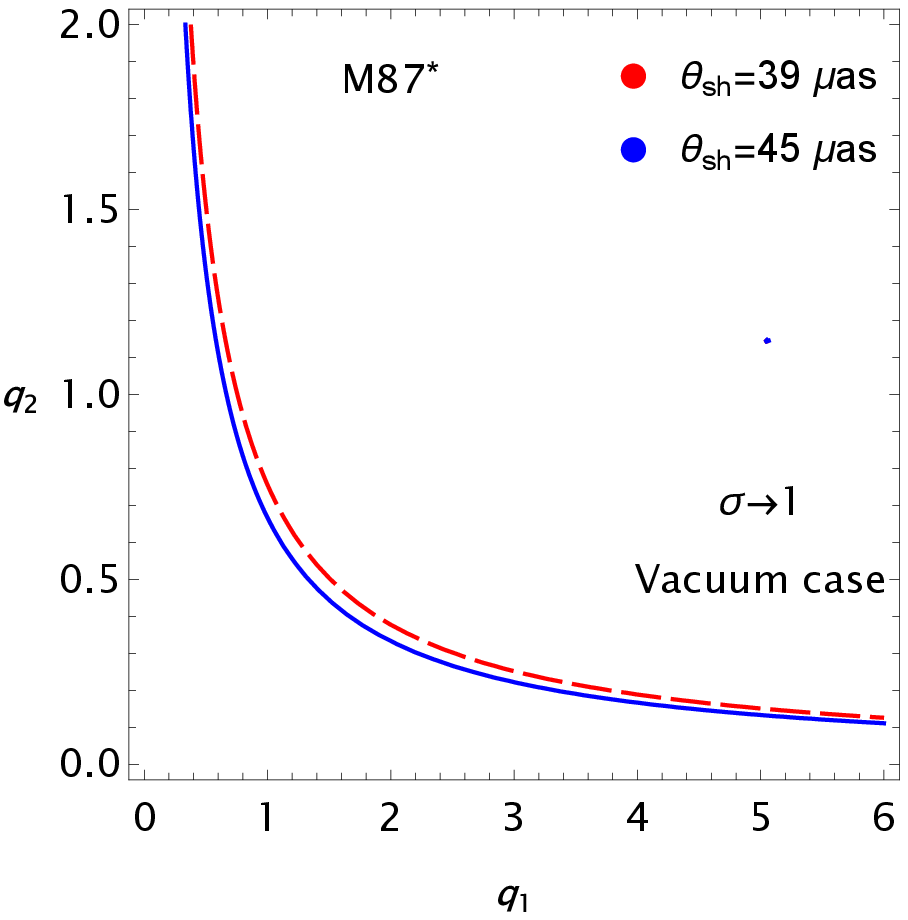}    \includegraphics[scale=0.8]{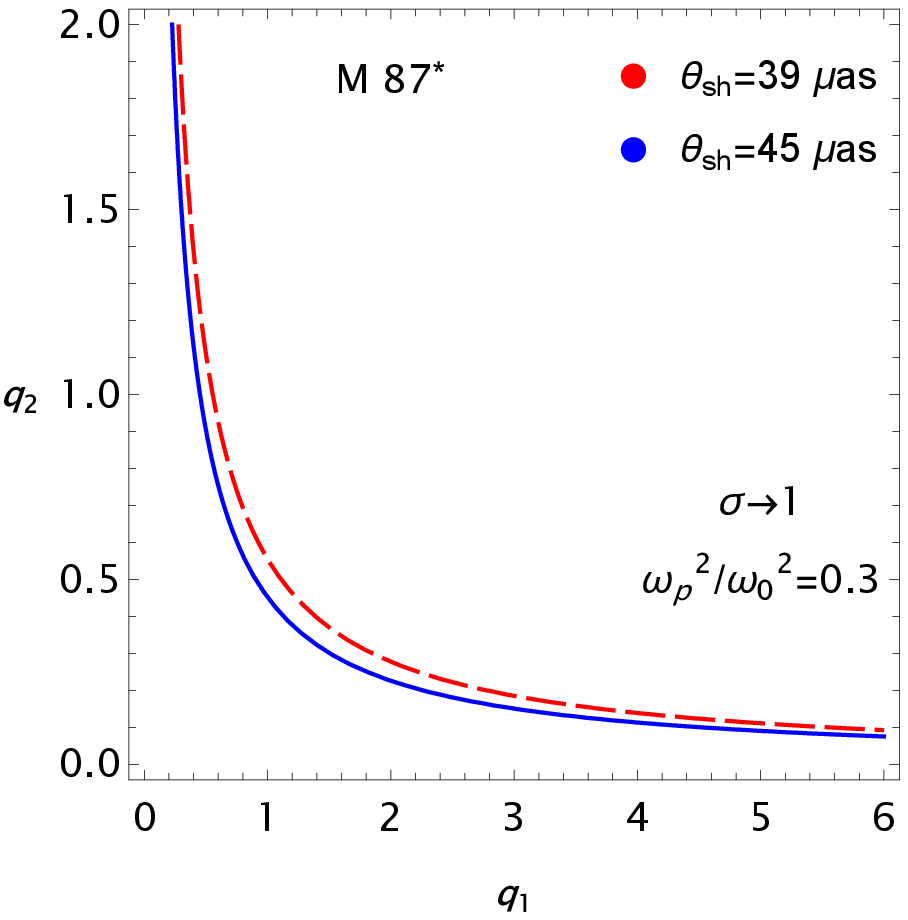}  \end{center} \caption{The same as Fig. \ref{plot:M87SgrA} but for $\sigma \to 1$.}\label{plot:M87SgrA1} \end{figure*}

In Figs. \ref{plot:M87SgrA} and \ref{plot:M87SgrA1}, depicted are the constraints on  possible values of the parameters $q_1$ and $q_2$ (between red-dashed and blue solid curves) using the observational angular size of SMBHs M87* and Sgr A* for $\sigma=0.5$ and $\sigma \to 1$. It is seen from the figure that the constrained range increases slightly  with the increase in the angular size. The ranges of $q_1$ and $q_2$ considerably increase in the  presence of a uniformly distributed plasma with the frequency ratio $\omega_p/\omega_0=\sqrt{0.3}$ (corresponding to a plasma frequency of about 127 GHz). In the limit $\sigma\to 0$, constrained values of $q_1$ and $q_2$ obey power-law behavior: $q_1 q_2= {\rm const}$, where the constant depends on the plasma frequency and the angular size. It is seen from  both figures that the value of the constant decreases as the plasma frequency and the angular size increase. However, in $\sigma=0.5$ case, $q_1$ and $q_2$ parameters have a complicated relationship depending on the size and plasma frequency. Moreover, in this case,  the constrained ranges of $q_1$ and $q_2$ parameters for M87* are bigger than that which for Sgr A*, due to its large mass.

\section{Discussions and Conclusion}\label{sec6}

In this paper, we have performed a systematic study
of the EMPG model \cite{dp-yeni} in the AdS background in the linear ghost-free limit in which quadratic and higher-order curvature terms are all dropped. The model is in the class of linear, torsion-free, metric-Palatini gravity theories \cite{harko2012,Capozziello2013a,Capozziello2015}, with the extensions that  a term  quadratic in the antisymmetric part of the affine curvature \cite{Vitagliano2011,Vitagliano2013,Demir2020,dp-yeni} exists. We performed a detailed investigation of shadow and photon motion around this black hole for the main purposes of probing the properties of the geometric Proca field. (As mentioned before, we use ``geometric Proca" to distinguish the massive non-metricity vector of the metric-Palatini gravity from generic Einstein-Proca systems as well as the $Z^\prime$ gauge boson literature.) 

The results obtained in this report can be summarized and discussed as follows:
\begin{itemize}

 \item 
 We have shown that the metric-Palatini gravity theory  in the AdS background reduces to the GR plus a geometrical massive vector theory, which we call the EMPG model \cite{Demir2020,dp-yeni}. 
 
 \item The EMPG system provides a novel geometro-dynamical framework. It can set the stage for diversely different physical phenomena. It would certainly have effects in various astrophysical media such as neutron stars, BHs, magnetars, and other compact objects. In the present work, we study BHs with asymptotically AdS solutions which are determined by solving the EMPG field equations (\ref{Einstein-eqns}) and (\ref{eom-Y}).

 \item Using the horizon structure of a gravitational compact object, the range of EMPG gravity parameters are also obtained in detail. It is shown that orbits enlarge or shrink depending on the value of the EMPG parameter $\sigma$. However, the effects of parameter $q_1$ and $q_2$ are opposite to each other. 

 \item We study the singularity structure of the EMPG black hole by analyzing the Kretschamann scalar (\ref{Kretsch-scalar}) and the Ricci-squared scalar (\ref{Ricci-scalar}). Since $q_1\rightarrow 0$ leads to the Schwarzschild solution, we simplify KS and RS expressions by expanding $q_1$ around zero up to the second order. The physical properties of the KS is studied and its behaviour is shown in Fig. \ref{plot:KS} for various parameter values $q_1$ at fixed values of $q_2$ and $\sigma$. We show that there are basically two singularities of the KS at $r=0$ and at the horizon $r = r_H$. One can see from  Fig. \ref{plot:KS} that the place of the singularity at the horizon depends on the model parameters and changes accordingly. Near $q_1=0$, as expected, the singularities in (\ref{Kretsch-scalar}) occur at $r=0$ and $r=r_H=2$, corresponding to the event horizon for the Schwarzschild black hole.  It is exactly seen from the KS (\ref{Kretsch-scalar}) that the Proca field causes the KS to deviate from the Schwarzschild limit by its charge ($q_1 \neq 0$ and $q_2 \neq 0$ ). The KS in the EMPG for $q_1 = q_2 \neq 0$ with $\sigma$ namely $0 \le \sigma < 1$ (ensuring $M_Y^2 \neq 0$) implies a Proca "hair" of the black hole. Given the singularity at the horizon the black hole is expected to have a ``hair", and that hair is provided by the Proca field.

\item  We show that the EMPG model is devoid of any metric instabilities since for our model the stability related parameter is zero which makes the Lyapunov exponent imaginary. The imaginary Lyapunov exponent accordingly prevents the perturbation of the metric to diverge. The critical impact parameter $b_c$ and the critical photon-trajectory depend on the time varying horizon radius which is dependent on the imaginary Lyapunov exponent so that for the EMPG model one expects no instability in the photon trajectory. The photon radius shows small variations about the EMPG horizon radius $r_H(0)$. Moreover, since in the EMPG model the Proca field couples minimally, one expects no instabilities arising normally from non-minimal couplings of the Proca field.
 
 \item Additionally, we have analyzed photon orbits and the influence of parameter $\sigma$. Photon sphere radius increases with $\sigma$ when $q_1$ or $q_2$ is fixed. Moreover, plasma effects on photon orbits have been studied for uniform and non-uniform cases.  In both cases, the photon sphere grows as the plasma frequency grows, but this effect is stronger in the uniform case.
 
 \item  We have discussed the observable quantity which is the shadow of a black hole studied by the EMPG spacetime metric in vacuum and plasma. In the vacuum case, increasing the value of $q_{1}$ or $q_{2}$, when one of them held fixed, decreases the shadow size. In the presence of plasma, the shadow size decreases as the plasma frequency increases, but we have seen that, for the same values of parameters, the photon sphere grows as plasma frequency grows. In other words, plasma brings photon sphere and shadow close to each other.

 \item {All the results (photon orbits, BH shadow) obtained above are compared to the case of usual Schwarzschild black hole spacetime, which corresponds to case $q_1 \to 0$, parametrically. It has been shown that under the effects of the parameter $q_1$ the size of BH shadow gets smaller.}

 \item Finally, we have determined the constraints on possible values of the parameters $q_1$ and $q_2$ using the observed shadow sizes of M87* and Sgr A* for $\sigma=0.5$ and $\sigma\to 1$. It is shown that the constraining ranges of the parameters increase with the increase of shadow size and plasma frequency.
\end{itemize}

\section*{Acknowledgements}

This research is partly supported by Grant F-FA-2021-510 of the Uzbekistan Ministry for Innovative Development.  D. D. and B. P.  acknowledge the contribution of the COST Action CA21106 - COSMIC WISPers in the Dark Universe: Theory, astrophysics, and experiments (CosmicWISPers). B. P. also acknowledges the contribution of the COST Action CA18108 - Quantum gravity phenomenology in the multi-messenger approach (QG-MM). The work of B.P. is supported by T{\"U}B{\.I}TAK B{\.I}DEB-2218 national postdoctoral fellowship program.

\bibliographystyle{apsrev4-1}  

\end{document}